\documentclass[english,aps,superscriptaddress,twocolumn]{revtex4}

\pdfoutput=1

\usepackage[T1]{fontenc}

\usepackage{graphicx}
\usepackage{epstopdf}
\usepackage{amssymb}
\usepackage{amsmath}
\usepackage{subfigure}
\usepackage[urlcolor=blue,hyperindex,colorlinks,bookmarks=true,linkcolor=black,citecolor=black]{hyperref}
\usepackage[normalem]{ulem}

 \newcommand{\be}{\begin{equation}}
\newcommand{\ee}{\end{equation}}
\newcommand{\bea}{\begin{eqnarray}}
\newcommand{\eea}{\end{eqnarray}}
\def\iden{\mathbb{I}}
\newcommand{\ket}[1]{\left|#1\right\rangle}
\newcommand{\bra}[1]{\left\langle#1\right|}

\newcommand{\abs}[1]{\lvert#1\rvert}
\newcommand{\kk}{\kappa}
\newcommand{\yy}{\gamma}

\begin{document}

\title{Theory of Josephson Photomultipliers: Optimal Working Conditions and Back Action}

\author{Luke C.G. Govia}
\affiliation{Institute for Quantum Computing and Department of Physics and Astronomy, University of Waterloo, Ontario, Canada}
\author{Emily J. Pritchett}
\affiliation{Theoretical Physics, Universit\"{a}t des Saarlandes, Saarbr\"{u}cken, Germany}
\author{Seth T. Merkel}
\thanks{present address: IBM T.J. Watson Research Center, Yorktown Heights, New York 10598, USA}
\affiliation{Theoretical Physics, Universit\"{a}t des Saarlandes, Saarbr\"{u}cken, Germany}
\author{Deanna Pineau}
\affiliation{Department of Physics and Astronomy, University of Victoria, British Columbia, Canada}
\author{Frank K. Wilhelm}
\affiliation{Theoretical Physics, Universit\"{a}t des Saarlandes, Saarbr\"{u}cken, Germany}
\affiliation{Institute for Quantum Computing and Department of Physics and Astronomy, University of Waterloo, Ontario, Canada}

\begin{abstract}
We describe the back action of microwave-photon detection via a Josephson photomultiplier (JPM), a superconducting qubit coupled strongly to a high-quality microwave cavity.  The back action operator depends qualitatively on the duration of the measurement interval, resembling the regular photon annihilation operator at short interaction times and approaching a variant of the photon subtraction operator at long times.  The optimal operating conditions of the JPM differ from those considered optimal for processing and storing of quantum information, in that a short $T_2$ of the JPM suppresses the cavity dephasing incurred during measurement. Understanding this back action opens the possibility to perform multiple JPM measurements on the same state, hence performing efficient state tomography. 
\end{abstract}
\maketitle

\section{Introduction}

Recently, many of the benchmark experiments of cavity quantum electrodynamics (QED) \cite{Haroche89,Rempe00,Haroche06,Kimble08} have been reproduced with superconducting circuits  \cite{Schuster07,Houck07,Leek07,Astafiev07,Fragner08,Grajcar08,Baur09,Lang11}, which operate in the quantum regime via exchange of microwave-frequency excitations \cite{Makhlin01,Shumeiko06,Nato06I,Insight,You05b}.  On these circuits, nonlinear devices couple to the microwave-frequency modes of transmission lines via ordinary circuit devices such as capacitors or inductors \cite{Blais04,Wallraff04,Schoelkopf08}, much as atoms couple to modes of a resonant electromagnetic cavity.   A fixed number of these artificial atoms can be fabricated in a given circuit, and their energy levels and interactions are tunable both at fabrication and during the course of an experiment.  For these reasons, circuit-QED (cQED) receives attention both as a possible platform for scalable, universal quantum computing \cite{Helmer09,DiVincenzo09} and for its ability to operate in regimes inaccessible by atomic cavity-QED \cite{Devoret07,Hofheinz09,Bourassa09,Casanova10,FornDiaz10,Niemczyk10,Wilson11,PRBR033}.

While many of the tools available in cavity-QED are straightforward to reproduce in the circuit analogue, the detection of single microwave-frequency photons proves challenging. Traditionally, the lower cutoff frequency of photon counters is determined by the work function or band gap of a certain material, which is at a minimum in the infrared range for stable materials.  
There are currently a few theoretical proposals for the construction of microwave photon counters \cite{Osberg09,Romero:2009fk,Romero:2009uq},  and recently it was demonstrated experimentally that a current-biased Josephson junction \cite{Caldeira81,Caldeira83,CA,CBA,Martinis85,Martinis87} can be used to count microwave photons \cite{Chen:2011ys}.  We refer to such a device as a Josephson photomultiplier (JPM), distinguishing it from a phase qubit \cite{Martinis02,Martinis09c} because the optimal operating conditions for photon detection are different than those required for storage of quantum information.  

Photon counters should be contrasted with amplifiers. While the former are sensitive to the intensity of the incoming radiation but not to the phase, the latter amplify the quadratures of the signal. Even though commercial microwave amplifiers operate far from the quantum limit, researchers have recently demonstrated single-photon sensitivity in  phase-preserving microwave amplification \cite{Mueck01,Mueck03,Hover11,Ribelli11,DeFeo10,Bergeal10,Castellanos08,Hatridge11,Clerk10}.   
By contrast, phase-insensitive photon counters have proven useful in quantum optics for reconstruction of the quantum state of light as, for example, in homodyne tomography. 
Workarounds to microwave photon counting based on linear amplification have been formulated \cite{Mariantoni05,Silva10} and demonstrated \cite{Menzel10,Bozyigit11,Mallet11}.  

As microwave photon counters, JPMs have an important application in efficient quantum state tomography of microwave photon states. 
Given that measurement by a JPM provides only limited information -- a click indicates the presence of one or more photons -- the post-measurement state still contains coherent information about the initial state. Following the idea of quantum regression  
this post-measurement state is connected to the pre-measurement state
by back action operators. Hence, if the back action operators are known, repeated measurements on a chain of post-measurement states can provide additional information about the original state.

In this paper, we theoretically model the back action of the proposed JPM, obtaining the precise relation between pre- and post-measurement states. This knowledge may allow efficient state tomography \cite{Paris04,Lvovsky09} including, for example, adaptive techniques \cite{Sarovar07}, or any other application requiring knowledge of the post-measurement state.  
We note that our results are based on a very abstract model and thus extend to other detection schemes whereby a quantum two-level system strongly couples to a resonant linear oscillator, so long as the observable detection event involves incoherent tunneling from an energy level of the two-level system (and not the oscillator). For example, this situation applies to some setups in atomic cavity QED.  We include realistic estimations of the energy dissipation and dephasing rates of a JPM, showing that operating the JPM in the regime of fast dephasing (short $T_2$) reduces the amount of dephasing incurred during measurement, and is thus advantageous.

In the following section, we discuss our model for the JPM.  In section III, we discuss the formalism of process tomography used to characterize the back action of the JPM.  In section IV, we give the back action both numerically and analytically in a variety of instructive and/or experimentally relevant regimes.  In section V, we discuss how to extract the operating regime of the JPM by simple tests with coherent light.  Finally, in section VI, we discuss the optimal working conditions of the JPM for the purpose of cavity state reconstruction.  

\section{Physical Model}
\begin{figure}[h]
\includegraphics[width = 0.9\columnwidth]{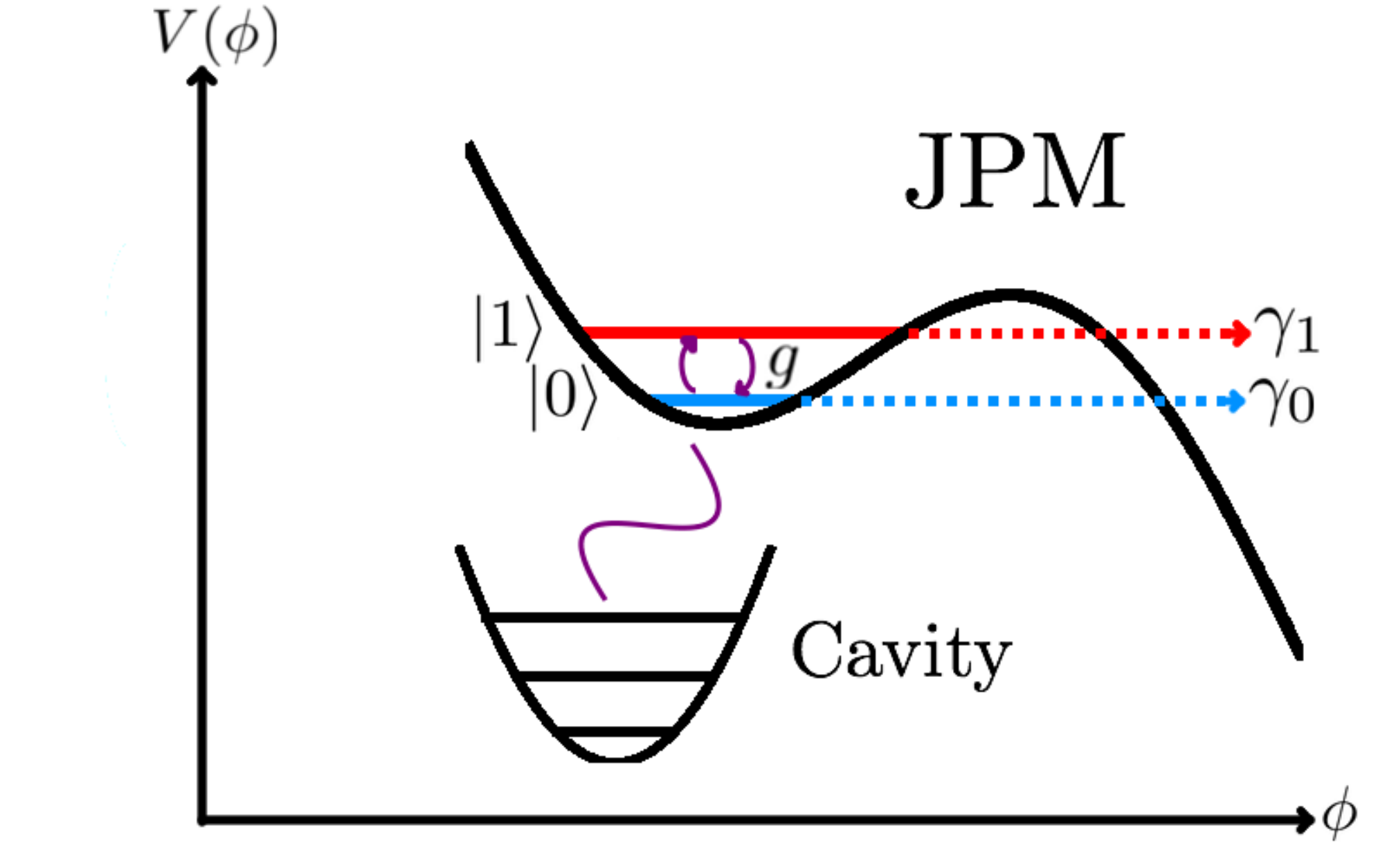}
\caption{This figure is a diagrammatic representation of the potential energy of a JPM as a function of the superconducting phase difference, and shows the interaction with a microwave cavity.}
\label{fig:JPMPot}
\end{figure}

A JPM consists of a current-biased Josephson junction (CBJJ)  \cite{Caldeira81,Caldeira83,CA,CBA,Martinis85,Martinis87} capacitively coupled to the microwave cavity of interest. 
The potential energy  of a JPM ,  shown in figure \ref{fig:JPMPot},  is 
\be
U(\phi)  = - I_c \frac{\Phi_0}{2 \pi} \cos{\phi} - I_b \frac{\Phi_0}{2 \pi}\phi
\ee
where $I_c$ is the critical current of the junction externally biased by current $I_b$, $\phi$ is the superconducting phase difference across the junction, and $\Phi_0 = \frac{\hbar}{2e}$ is the magnetic flux quantum.
We consider a junction biased such that the potential well contains only a few meta-stable states. All of these states can tunnel incoherently out of the potential well, but due to the exponential relationship between tunnelling rate and barrier height, the tunnelling rate for higher energy states is several orders of magnitude higher than that for the ground state. Note that, in analogy to the phase qubit, the current source can be replaced by a large, flux-biased superconducting loop \cite{Simmonds04,Martinis09c}.

Photon detection relies on an incident microwave photon to transition the JPM to its first excited state. This transition is enhanced by pulsing the bias current, bringing the energy level splitting of the ground and first excited state of the JPM on- or near-resonant with the microwave cavity.  Once the JPM reaches its excited state, it tunnels more rapidly out of the metastable state.  This tunneling process is incoherent, resulting in a measurable voltage pulse in the circuit that is interpreted as the detection of a single photon
\cite{CA,CBA,Martinis87,Martinis85,Chen:2011ys}. By comparison, a related method of cavity state reconstruction determines the number of cavity photons present by the frequency of coherent oscillation between the cavity and a phase qubit [\onlinecite{Hofheinz09}].

We assume that the JPM and cavity states are initially separable with the JPM in the lowest energy metastable state. Physically, this means that the JPM and microwave cavity must be brought on resonance adiabatically with respect to the JPM's internal evolution, but non-adiabatically with respect to the cavity-JPM interaction.
We describe the full system's Hilbert space with tensor products of single mode cavity eigenstates and three detector states, $\left\{ \ket{0}_{\rm d}, \ket{1}_{\rm d}, \ket{m}_{\rm d} \right\}$.  The states $\ket{0}_{\rm d}$ and $\ket{1}_{\rm d}$ correspond to the ground and first excited metastable states of the JPM, while the measured state $\ket{m}_{\rm d}$ is an amalgamation of the many possible states that the JPM can tunnel into incoherently, producing an observable output voltage.   
\label{text:preparation}

The coherent interaction between the cavity and the JPM as well as the relevant incoherent processes (for example, tunneling into the measured state, dephasing, and relaxation) are described by the quantum master equation
\bea
\label{eqn:masterequation}
&&\nonumber \dot{\xi}(t) = \hat{S}[\xi(t)] \\
&&= -i[\hat{H},\xi(t)] + \sum_{\mu}\left(\hat{J}_\mu\xi(t)\hat{J}_\mu^\dagger-\frac{1}{2}\{ \hat{J}_\mu^\dagger \hat{J}_\mu,\xi(t)\}\right)
\eea
where $\xi(t)$ is the cavity-JPM system's density matrix. Here $H$ is the Jaynes-Cummings interaction
\be
\label{eqn:Hamiltonian}
\hat{H} \equiv g(\hat{a}^\dagger \hat{\sigma}^- + \hat{a} \hat{\sigma}^+)
\ee
where $\hat{a}$ and $\hat{a}^\dagger$ are the lowering and raising operator associated with the cavity mode, $\hat{\sigma}^{\pm}$ are the lowering and raising operators between the states $\ket{0}_{\rm d} $ and $\ket{1}_{\rm d} $, and  $g$ is the coupling strength between the cavity and the JPM.  Note that this Hamiltonian conserves total excitation number and does not couple photons coherently to the measured state, which simplifies the following analysis.

A set of Linblad operators $\{\hat{J}_\mu\}$ describe the relavent incoherent processes. 
\bea
\label{eqn:DC}
 \hat{J}_1\equiv\sqrt{\gamma_1}(\mathbb{\hat{I}}_{\rm c}\otimes\ket{\rm m}\bra{1}_{\rm d})
 \eea
 describes incoherent tunneling out of the excited metastable state 
leaving one less excitation in the cavity, and thus corresponding to photon measurement.  Tunneling out of the metastable ground state  is described by
  \bea
\label{eqn:DC}
 \hat{J}_0\equiv\sqrt{\gamma_0}(\mathbb{\hat{I}}_{\rm c}\otimes\ket{\rm m}\bra{0}_{\rm d})
 \eea
where $\gamma_0$ is the effective dark count rate since a measurement signal is produced without changing the number of excitations in the cavity.   We also include Linblad operators to describe pure dephasing of the JPM over characteristic time $T_2$
\be
\label{eqn:dec}
\hat{J}_2\equiv\frac{1}{\sqrt{T_2}}(\mathbb{\hat{I}}_{\rm c} \otimes \ket{1}\bra{1}_{\rm d})
\ee
and  energy relaxation from the excited state to the ground state of the JPM over characteristic time $T_1$
\be
\label{eqn:rel}
\hat{J}_3\equiv \frac{1}{\sqrt{T_1}}(\mathbb{\hat{I}}_{\rm c}\otimes\ket{0}\bra{1}_{\rm d}).
\ee
In general the cavity decoheres as well \cite{Goeppl08,Megrant12}, but this happens slowly compared to other relevant timescales.

\section{Process Tomography}

For tomography of the process of cavity state measurement by a JPM, we calculate the Liouville supermatrix ${\cal T}(t)$ generated by 
\begin{eqnarray}
&&\nonumber{\cal S} \equiv -i\left( \hat{H} \otimes \hat{\iden} - \hat{\iden} \otimes \hat{H} \right) \\
&&\nonumber+ \sum_{\mu} \left( \hat{J}_\mu \otimes \hat{J}_\mu  - \frac{1}{2} \hat{J}_\mu^\dagger \hat{J}_\mu \otimes \hat{\iden} - \frac{1}{2}\hat{\iden} \otimes \hat{J}_\mu^\dagger \hat{J}_\mu \right),
\end{eqnarray}
which transforms an arbitrary, vectorized initial cavity-JPM state to the solution of (\ref{eqn:masterequation}):
\be
\vec{\xi}(t) = {\cal T}(t)\vec{\xi}(0)= e^{{\cal S}t} \vec{\xi}(0).
\ee
We then transform the Liouville supermatrix into the more commonly used $\chi$-matrix representation \cite{NielsenChuang,Mohseni:2008fk}, which, for a given basis $\{\hat{E}_{\mu}\}_{\mu=0}^{N^2-1}$ of operator space ${\cal L}({\cal H})$,
satisfies
\be
\xi(t) = \sum_{\mu\nu = 0}^{N^2 - 1} \chi_{\mu\nu}(t) \hat{E}_{\mu} \xi(0) \hat{E}_{\nu}^\dagger.
\ee
\noindent 
In the standard basis, $E_{\mu(\alpha,\beta)} \equiv \ket{\alpha} \bra{\beta}$ where $\mu(\alpha,\beta) \equiv (N \times \alpha) + \beta$ and $\{\ket{\alpha}\}_{\alpha=0}^{N-1}$ is an eigenbasis of the full system's noninteracting Hamiltonian, the $\chi$-matrix elements are simply
\be
\chi_{\mu(\alpha,\beta)\nu(\gamma,\delta)}(t) \equiv \chi_{\alpha\beta\gamma\delta}(t) = \bra{\alpha}  \left(e^{{\cal S}t}\ket{\beta} \bra{\delta}\right) \ket{\gamma}
\ee 
and can be obtained by a permutation of the Liouville supermatrix elements:
$\chi_{\alpha\beta\gamma\delta} = T_{\alpha\gamma\beta\delta}$.
In all calculations, we assume an initial state of the form
\begin{equation}
\xi(0)=\rho_{\rm c}(0)\otimes|0\big>\big<0|_{\rm d},
\end{equation}
a product state of the initial cavity state, $\rho_{\rm c}(0)$, and the lowest energy metastable state of the JPM. Preparation of this factorized state has already been described earlier (section {\ref{text:preparation}}).

We are interested in the back action of the JPM onto the cavity state conditioned on measurement outcome $s\in\{0,1,{\rm m}\}$,
\begin{equation}
\label{eqn:ReducedDensityMarix}
\rho^s_{\rm c}(t)=\frac{\big<s|e^{\hat{S}t}\xi(0)|s\big>_{\rm d}}{P^s(t)},
\end{equation}
where $P^s(t)$ normalizes the cavity state by the probability of obtaining JPM final state $|s\big>_{\rm d}$. Only incoherent tunneling into the measurement state is allowed by our model, so no coherent superposition between the measured and non-measured state is possible. Therefore, $\rho_{\rm c}^{s={\rm m}}$ gives the cavity state after detection of a photon, and in the case that no photon is detected, the cavity will be in a mixture of the states with $s=0,1$.
Each outcome is completely described by an off-diagonal  $d_{{\rm cav}}^2 \times d_{{\rm cav}}^2$ block of the full $\chi$-matrix, which is by itself a valid $\chi$-matrix of the isolated cavity.  We label these reduced $\chi$-matrices  $\chi^s$, which completely describe the evolution of an arbitrary initial state $\rho_{\rm c}(0)$, in the case of measurement outcome $s$.  

\section{Analytical and Numerical Solutions for the $\chi$-Matrix}
\subsection{No Tunneling Model\label{sec:notunnel}}
\label{sec:AnChi}

To understand the back action of photon detection, we first consider a simpler model where there are no incoherent processes, and measurement of a photon corresponds to projecting the JPM onto its metastable excited state $|1\big>_{\rm d}$ (rather than $|{\rm m}\big>_{\rm d}$).  
In this model, eq. (\ref{eqn:masterequation}) reduces to
$
 \label{eqn:simplemasterequation}
\dot{\xi}(t) = -i[\hat{H},\xi(t)] 
$, 
and the cavity conditioned on measurement outcome $s$ at time $t$ can be expressed in terms of a single time-dependent back action operator $\hat{B}^s(t)$ acting only on the Hilbert space of the cavity:
\begin{equation}
\rho^s_{\rm c}(t)=\frac{\hat{B}^s(t)\xi(0)\hat{B}^{s} (t)^\dagger}{P^s(t)}.
\end{equation}
Furthermore, in this model, the $\hat{B}^s(t)$ are straightforward to calculate explicitly:
\bea
\label{eqn:B1}
{\hat{B}}^1(t)&\equiv&\big<1|\xi(t)|0\big>_{\rm d}\\
&=&-i\sum_{n=1}^\infty\sin(gt \sqrt{n})\ket{n-1}\bra{n}_{\rm c} \;  \; \nonumber\\
{\hat{B}}^0(t)&\equiv&\big<0|\xi(t)|0\big>_{\rm d}=\sum_{n=0}^\infty\cos(gt\sqrt{n})\ket{n}\bra{n}_{\rm c}.
\eea
The cavity is initialized in a superposition of $n$-photon Fock states, and when only coherent cavity-JPM interaction is included, each $n$-photon Fock state in superposition will exchange a single excitation with the JPM at a Rabi frequency $g\sqrt{n}$.  Measurement projects the detector onto $|0\big>_{\rm d}$ or $|1\big>_{\rm d}$, modifying the cavity with back action $\hat{B}^0(t)$ or $\hat{B}^1(t)$, respectively. 
From these operators we obtain the average detection probability 
\bea
P^1(t)&=&\big<\hat{B}^{1\dagger}(t)\hat{B}^1(t)\big>_0\nonumber\\&=&\sum_nP^1_n(t)\bra{n}\rho_c(0)\ket{n}
\eea
where $P^1_n(t)\equiv \sin^2(gt\sqrt{n})$ is the detection probability when the $n$-photon Fock state is initially prepared.

  It is possible, by averaging over repeated measurements at increasing time intervals, to distinguish Fock states and incoherent mixtures of Fock states by Fourier transforming the average detection probability $P^1(t)$,
as was demonstrated in ref. \cite{Hofheinz08}; however, more sophisticated state tomography is required for resolving superpositions of Fock states.  One approach is to displace the cavity state and reconstruct a convenient phase space description of its initial state, repeating measurements and averaging at every point in phase space that is resolved \cite{Hofheinz09}.  Here we look for a quantum description of the measured cavity state so that repeated measurements on a single input state can be used for a more efficient state tomography.

From the behavior of  (\ref{eqn:B1}), 
\be
\label{eqn:B1Order}
\hat{B}^1 = (-igt)\hat{a} + {\cal O}(\sqrt{n}gt)^3,
\ee
we can see that at short-times (and finite $n$),  the measurement back action is proportional to the photon lowering operator.  Furthermore, at short times, different Fock states can be distinguished by their tunneling rate into the detector, $\Gamma_n\equiv g^2n$.  However, at longer times $t\gtrsim t_n\equiv1/g\sqrt{n}$, the oscillations from different Fock states become out of phase and difficult to distinguish.  This effect is described by the correction to the $\hat{a}$ operator in (\ref{eqn:B1Order}).

We note that back action is well-described by the lowering operator when the interaction times $t_n$ (for the largest occupied $n$) are the largest time scales of the system.   This happens, for example, in the free photon regime where coupling strengths are very weak \cite{Carmichael99}.
Also, we will later show that in the richer model where detection corresponds to tunneling out of the $\bra{1}_{\rm d}$ at the rate $\gamma_1$, the lowering operator is a good approximation to the back action for times shorter than the excited state tunneling time $\gamma_1^{-1}$. 
To detect low-energy microwave photons, however, we need measurement times long compared to $t_{\rm n}$ and $\gamma_1^{-1}$, and  corrections to the lowering operator become important to fully understand the back action associated with photon detection.

\subsection{Full (Tunneling) Model}
 
 In the full model, a detection event corresponds to the detector incoherently tunneling to the `measurement' state $|{\rm m}\big>_{\rm d}$, which corresponds to the JPM observably tunneling from a metastable state to a near continuum of levels.  The JPM does not reset to its initial state on a time scale comparable to the measurement interval and therefore can not resolve the total number of photons present in the cavity, only whether there is at least one present.  Therefore, we do not expect the measurement back action on the cavity to be exactly the photon annihilation operator, but rather an operator of the form
 \be
 \label{eqn:backaction}
\hat{B}_{\rm m}=\sum_{n=1}^\infty|n-1\big>\big<n|.  
\ee
which we refer to as the subtraction operator. We note that this back action can be used to separate the number-dependent part of the annihilation operator, 
 \be
\hat{a}\equiv \hat{B}_{\rm m} \hat{N}^{1/2},
\ee
and was thus considered as the exponential of  a quantum phase operator by Susskind and Glogower \cite{Susskind64}, but a currently more accepted unitary version was proposed by Pegg and Barnett \cite{Pegg89}. 

 While in the no-tunelling model the measurement back action reflects undamped Rabi oscillations between the cavity and JPM, we expect these oscillations to be damped by incoherent tunneling out of the metastable states of the JPM so that, when averaged over an entire measurement interval, the back action has the form of (\ref{eqn:backaction}).  Transitions occur from $|n\big>$ to  $|n-1\big>$ photon Fock states, with no preference on the number $n$ of photons originally present.  However, because the {\it initial} tunneling rate depends on the number of photons present, and it takes time for this averaging effect to occur, we expect that for measurement intervals short compared to $t_n$ and $\gamma_1^{-1}$ the back action will more closely resemble that of the usual photon-number resolving annihilation operator.
 
To understand the distinguishing signatures of $\hat{a}$ and $\hat{B}_{\rm m}$ in the framework of process tomography, it is instructive to examine the $\chi^1$ matrices corresponding to each.  Both will have 
\bea
\label{eqn:betadef}
\chi^1_{j-1jk-1k}&\equiv&\beta_{jk}\\
 &\forall& j,k\in\{1,..., N-1\}\nonumber
\eea
non-zero, 
corresponding to superpositions of $\ket{j}$ and $\ket{k}$ photons transitioning to $\ket{j-1}$ and $\ket{k-1}$ photons.  For a good photon detector, the number of excitation in a given Fock state is decreased by exactly one, therefore all other elements of $\chi^1$ are zero.   When the back action operator is $\hat{a}$, 
\be
\beta_{jk}=\sqrt{jk},
\ee
while for $\hat{B}_{\rm m}$, 
\be
\beta_{jk}=1
\ee
for all values of $j,k\in\{1,..., N-1\}$.  
In the following section, we numerically study the time-dependence of the $\beta_{jk}$ in our full physical model, using the values of $\beta_{jk}$ for known examples of back action models as a point of reference. 
 
\subsection{Numerical Simulations for the $\chi^1$ Matrix}
\subsubsection{Bare JPM}
\label{sec:IdJPM}
Here we present the $\chi^1$-matrix elements numerically generated using the Liouville supermatrix approach, first in the case of a bare detector experiencing no dark counts, dephasing, or energy dissipation ($\gamma_0=0$ and $T_1=T_2=\infty$).  In this case, the $\chi^1$-matrix has the same nonzero elements as those for  $\hat{a}$ and $\hat{B}_{\rm m}$, labelled above as $\beta_{jk}$.

\begin{figure}[h]
\subfigure[Bare JPM Diagonal Matrix Elements]{
\label{fig:BareDiag}
\includegraphics[width = \columnwidth]{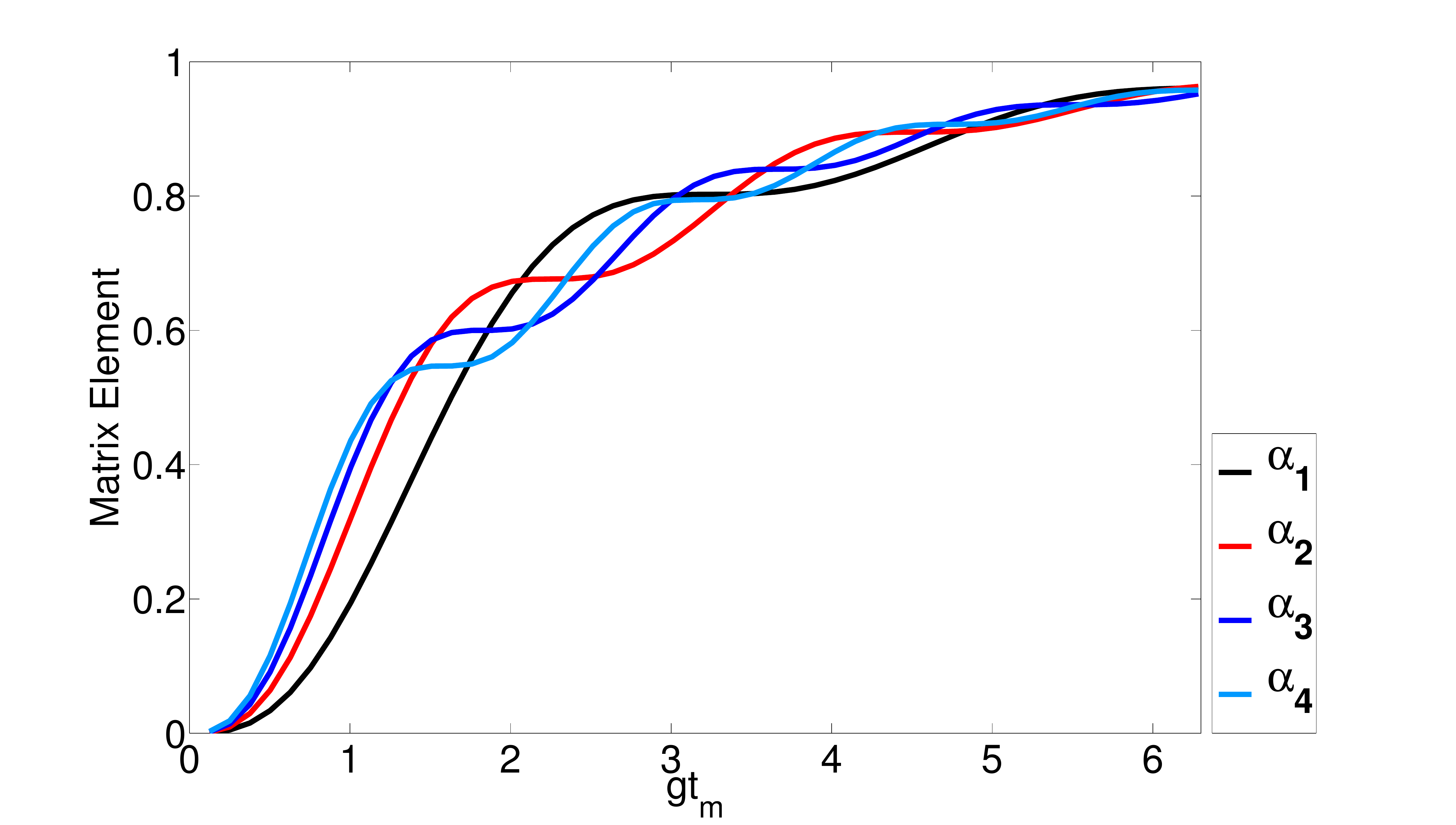}}
\subfigure[Bare JPM Off-Diagonal Matrix Elements]{
\label{fig:BareOff}
\includegraphics[width = \columnwidth]{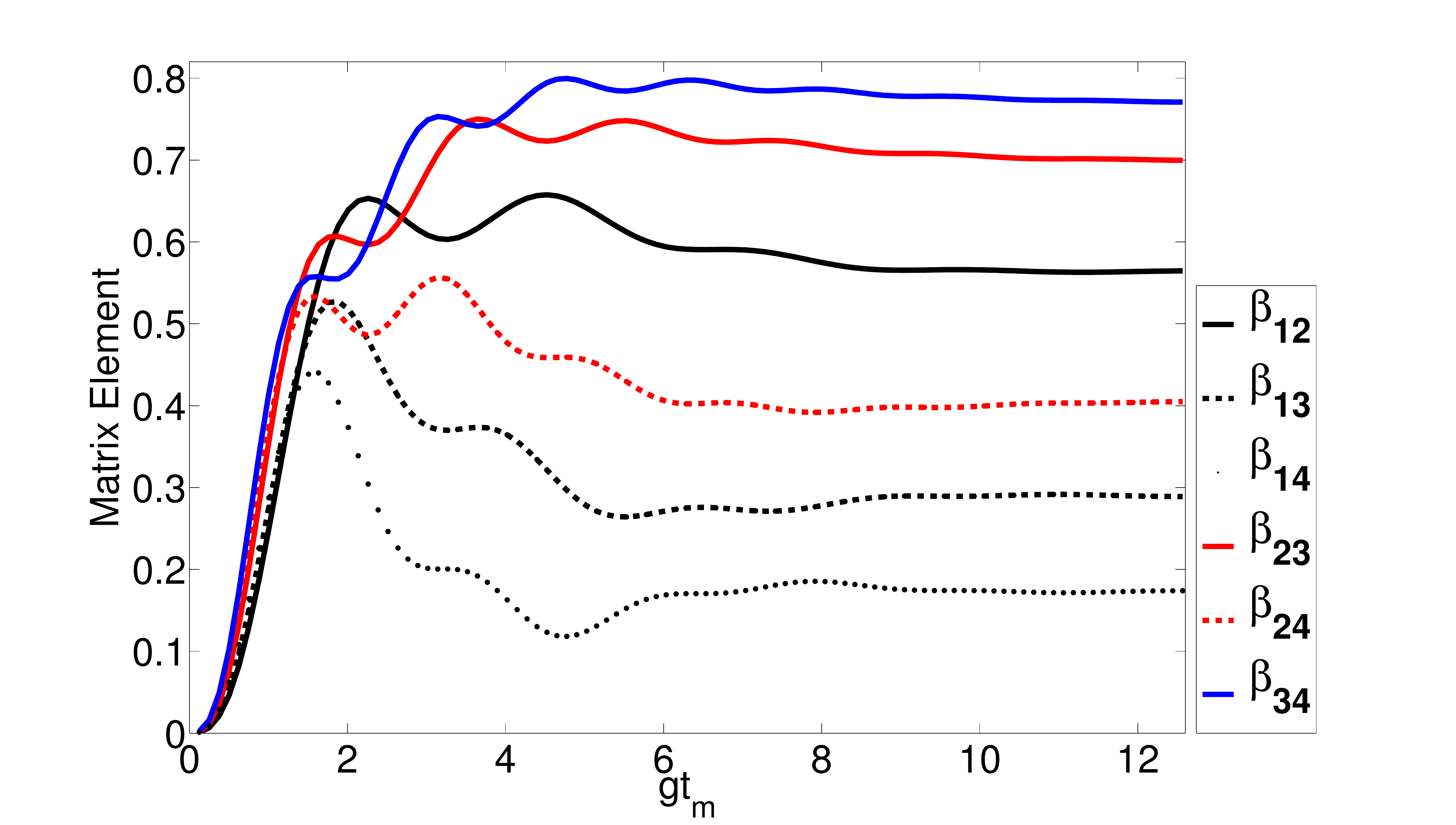}}
\caption{(a) Diagonal and (b) off-diagonal $\chi^1$-matrix  elements for a bare JPM as a function of time, where $\alpha_{j} = \beta_{jj}$ as defined in equation \ref{eqn:betadef}.}
\label{fig:Bare}
\end{figure}

The $\beta_{jk}$ are plotted as a function of total measurement time ($t_m$) in figure \ref{fig:Bare} for $j,k\in\{1,2,3,4\}$. 
As is demonstrated clearly by the diagonal $\alpha_j\equiv\beta_{jj}$ plotted in figure \ref{fig:BareDiag}, the $\chi^1$-matrix elements show oscillatory behaviour at $n$-dependent frequencies,  as in equation (\ref{eqn:B1}), $g\sqrt{n}$.  Similarly, the off-diagonal elements also show oscillatory behaviour, as can be seen in figure \ref{fig:BareOff}. 

In the long time limit, the diagonal elements all tend to unity as expected for a back action resembling the subtraction operator, however, the off-diagonal elements do not. This additional dephasing can be explained by the uncertainty in time of the switching event, which is of the order $\gamma_1^{-1}$. As the phase of the off-diagonal matrix elements precesses with frequencies proportional to $g$, this uncertainty gives a spread of the phase of size $g/\gamma_1$. We will later see how decoherence reduces this uncertainty, thus reducing the amount of dephasing incurred by measurement.

\subsubsection{Pure Dephasing}

\begin{figure}[h]
\subfigure[Bare JPM and Pure Dephasing Diagonal Matrix Elements]{
\label{fig:T2Diag}
\includegraphics[width = \columnwidth]{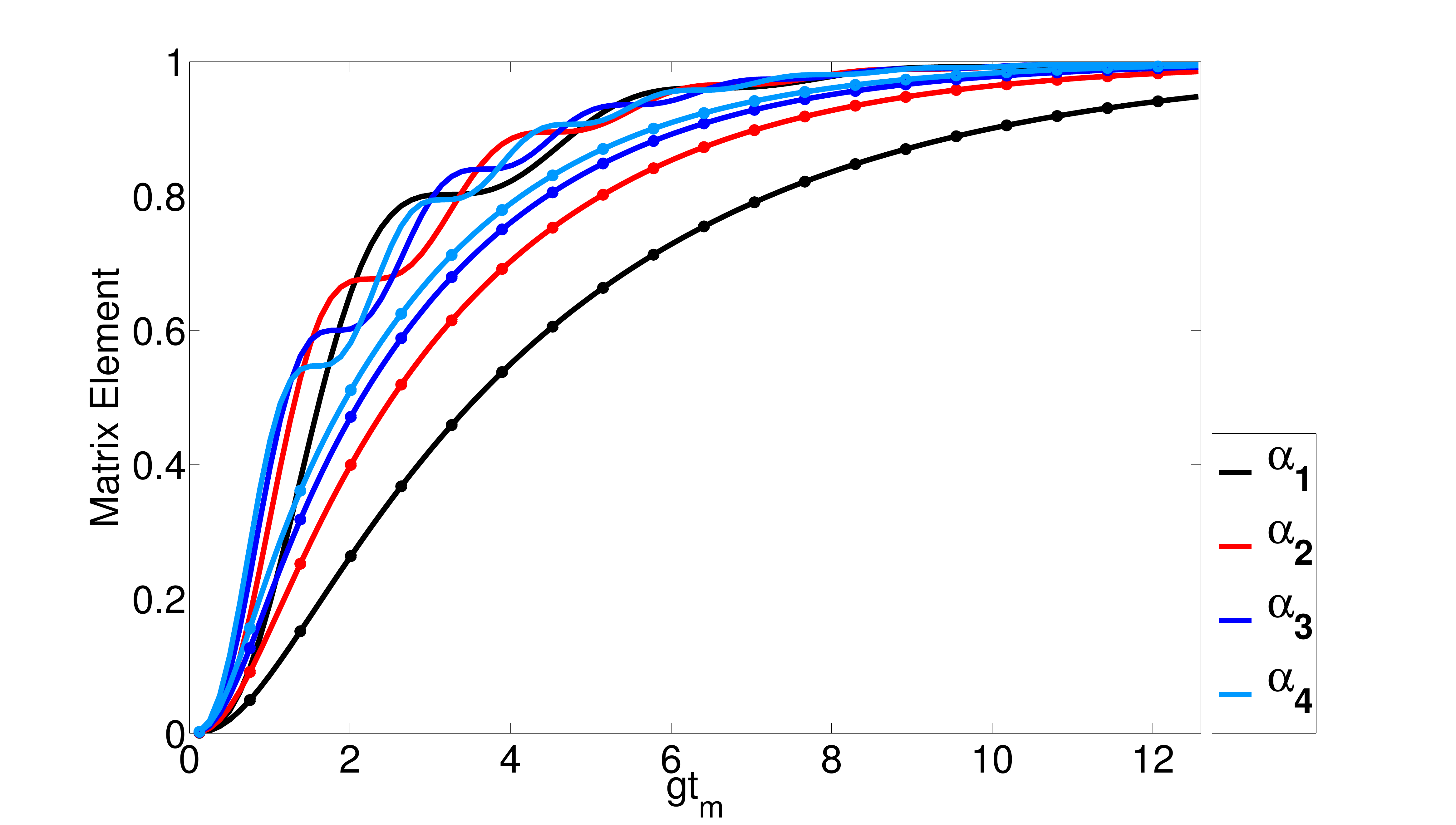}}
\subfigure[Bare JPM and Pure Dephasing Off-Diagonal Matrix Elements]{
\label{fig:T2Off}
\includegraphics[width = \columnwidth]{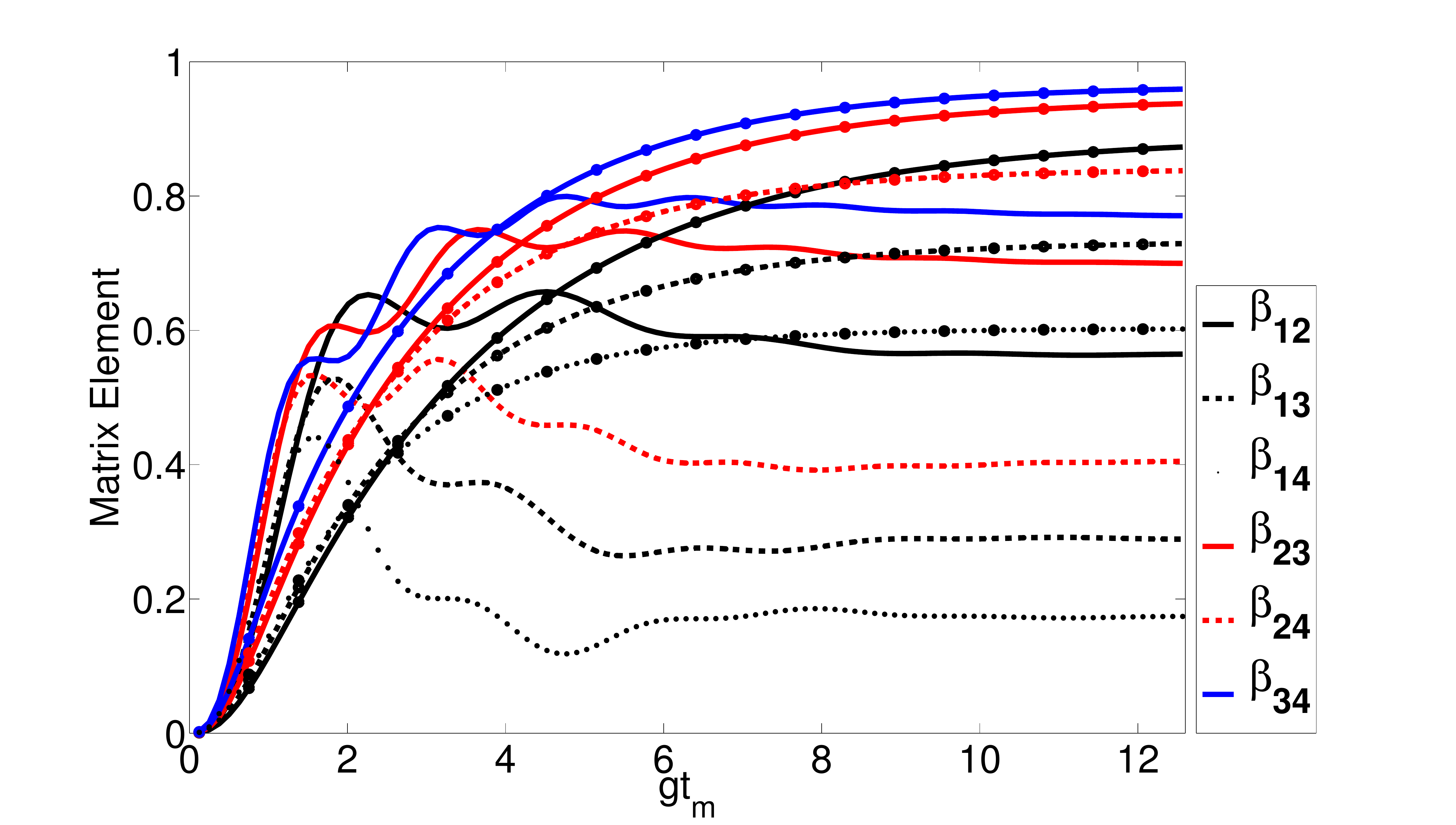}}
\caption{Comparison of the (a) diagonal and (b) off-diagonal $\chi^1$-matrix elements of a bare JPM with one experiencing additional pure dephasing $1/T_2=10\gamma_1$ (marked with circles).   In both plots, each curve's color indicates its row in the $\chi^1$-matrix, and in (b), distance from the diagonal is indicated by line style.}
\label{fig:T2Per}
\end{figure}

We now consider a JPM that experiences pure dephasing between its ground and excited states, as would be described by a master equation including the Lindblad operator  $\hat{J}_2$ of equation (\ref{eqn:dec}). In this case, the $\chi^1$-matrix has the same non-zero elements as that for the bare detector since the selection rules imposed by the conservation of excitation number are still valid. The bare detector and pure dephasing $\chi^1$-matrix elements are compared in figure \ref{fig:T2Per}, where the dephasing time has been chosen such that $\frac{1}{T_2} = 10\ \gamma_1$, deep in the strong dephasing regime.

As can be seen in figure \ref{fig:T2Per}, decreasing the value of $T_2$ has multiple effects. On the one hand, the photon transfer from the cavity into the detector is slowed down by decreasing $T_2$, so at short time the $\chi^1$-matrix elements are smaller. Also, the coherent oscillations are damped as $T_2$ has the effect of turning the coherent tunnelling between the cavity and the JPM into an incoherent process, \label{text:T2CO} similar to the crossover from strong coupling cavity QED to the Purcell regime \cite{Haroche06}. In fact, a phase Purcell effect has been discussed in \cite{Ioana07a,Ioana07b}.  This affects both the diagonal and off-diagonal $\chi^1$-matrix elements.

\begin{figure}[h]
\subfigure[Off-Diagonal Asymptotic Limits]{
\label{fig:T2Asymp}
\includegraphics[width = \columnwidth]{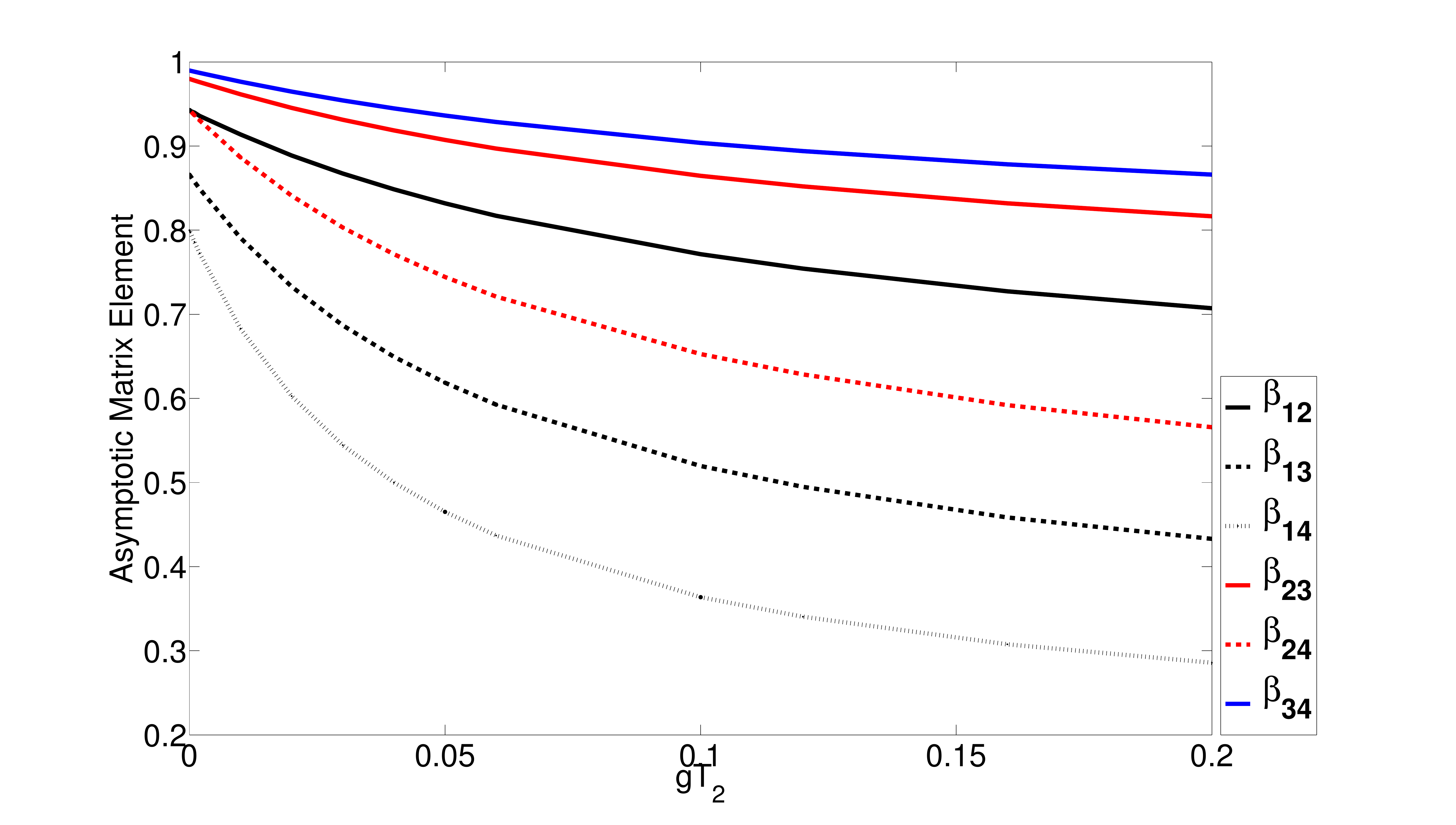}}
\subfigure[Off-Diagonal Asymptotic Time Scale]{
\label{fig:T2Times}
\includegraphics[width = \columnwidth]{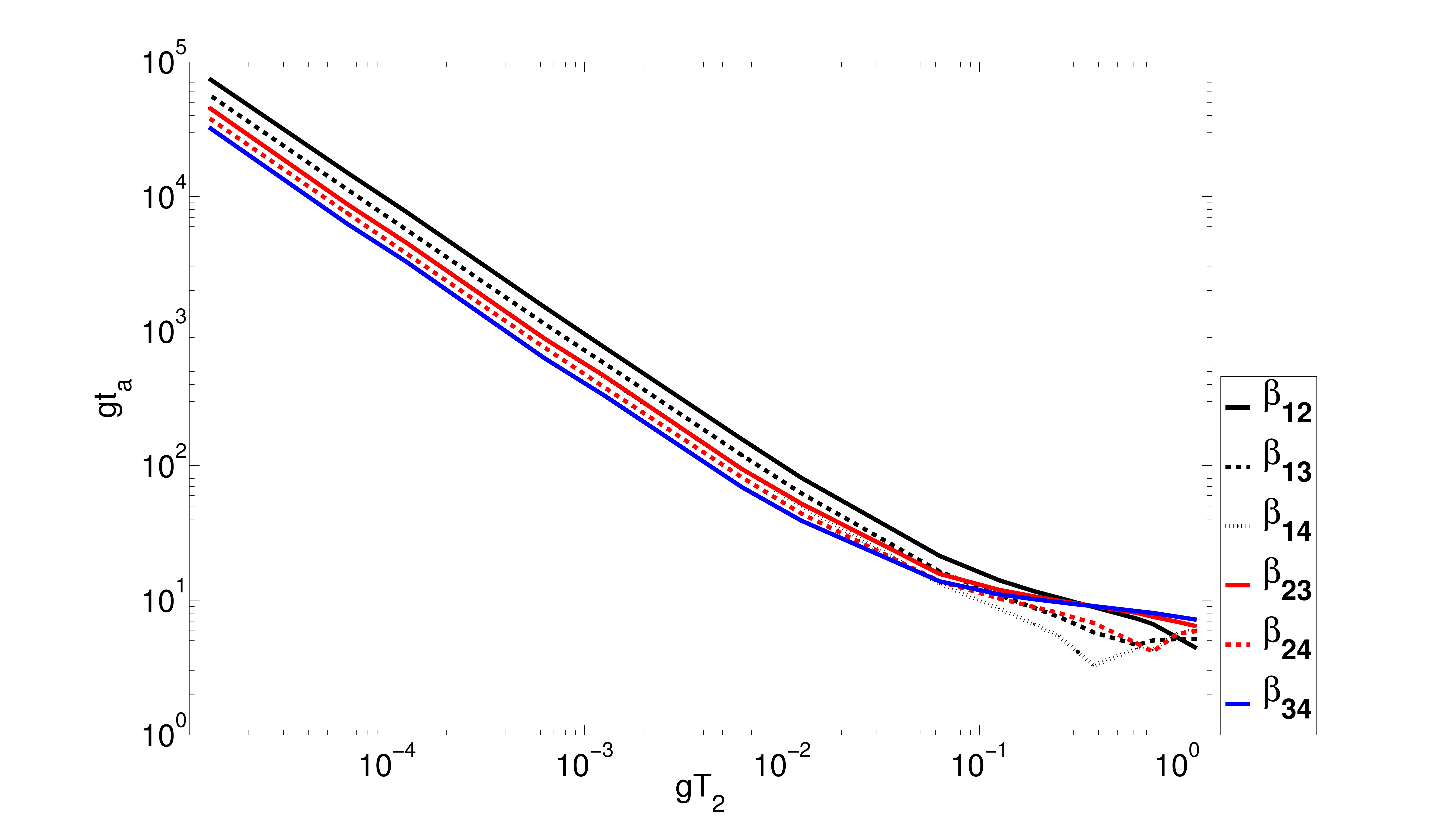}}
\caption{Figure (a) shows the asymptotic limit of the off-diagonal $\chi^1$ matrix elements as a function of $T_2$, and figure (b) shows the timescale over which these asymptotic limits are reached. Figure (b) is not monotonic due to the coherent oscillations present for long $T_2$.}
\label{fig:Bare}
\end{figure}

Once the photon transfer efficiency is no longer a limiting factor, the asymptotic limit of the diagonal $\chi^1$-matrix elements is not affected by $T_2$. The off-diagonal elements saturate to a value that is set by measurement-induced dephasing which is lowered by short $T_2$ (as seen in figure \ref{fig:T2Asymp}). This reduction in measurement-induced dephasing is due to the fact that $T_2$ turns the coherent tunnelling between cavity and JPM into an incoherent process, and thus reduces the phase precession of the off-diagonal elements and with it the uncertainty of these phases at the moment of measurement.   Although the amount of dephasing decreases with decreasing $T_2$, the total measurement time required for the $\chi$-matrix elements to reach their asymptotic value increases, as shown in figure \ref{fig:T2Times}.

\begin{figure}[h]
\subfigure[Coherent State Detection Probability]{
\label{fig:T2CDet}
\includegraphics[width = \columnwidth]{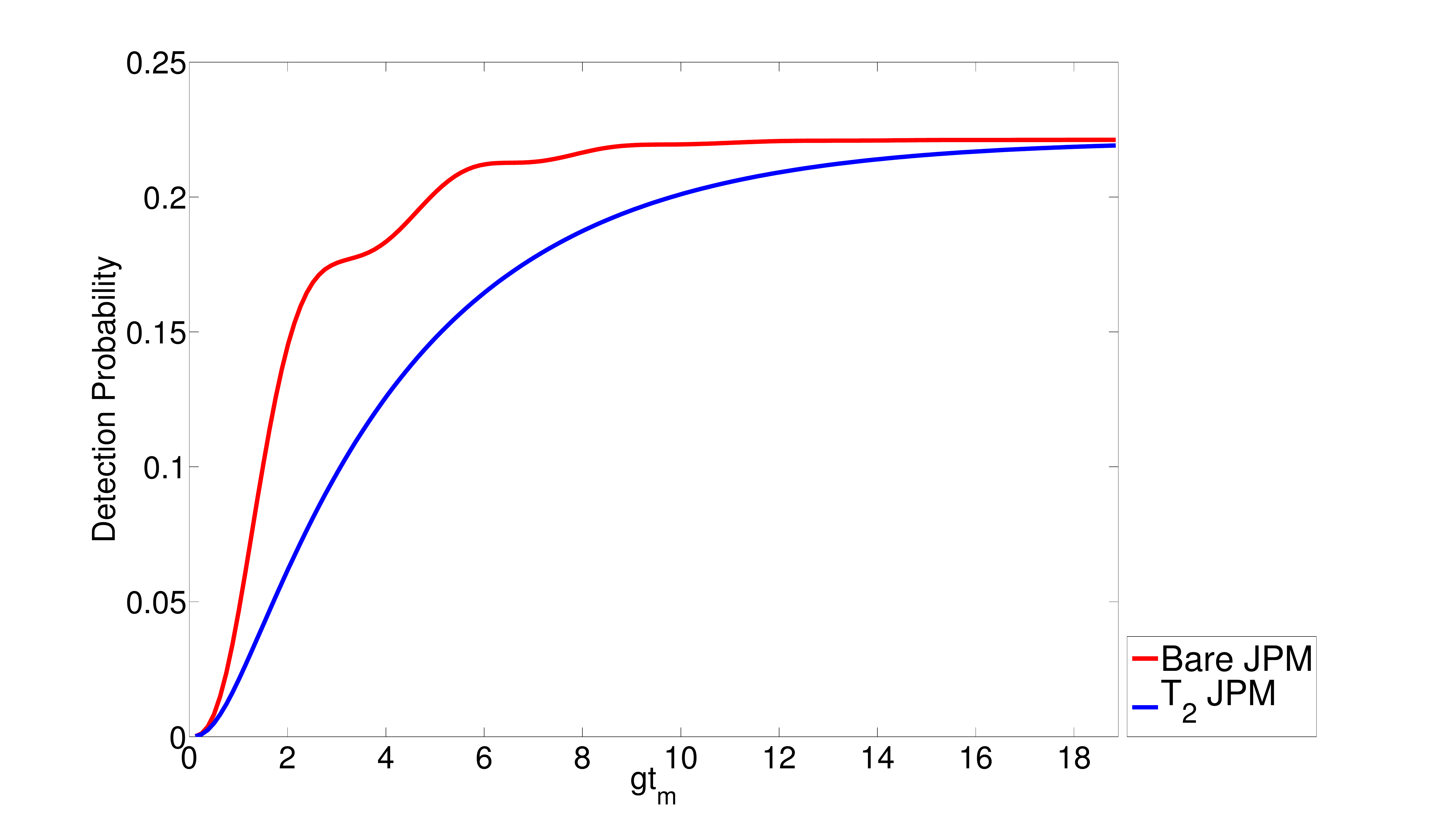}}
\subfigure[One Photon Fock State Detection Probability]{
\label{fig:T2FDet}
\includegraphics[width = \columnwidth]{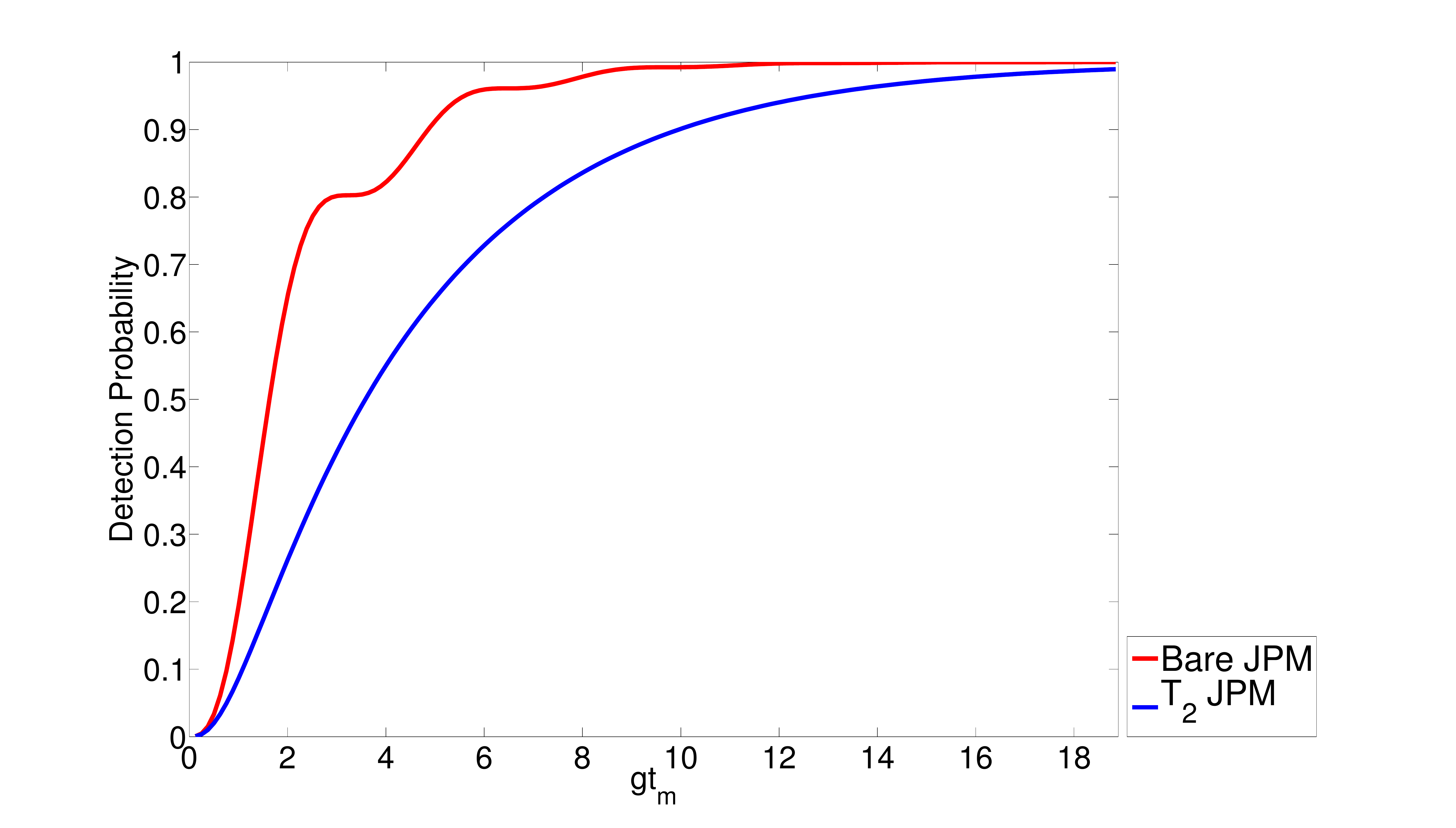}}
\caption{ Figure (a) shows the detection probability for a bare JPM and a JPM experiencing pure dephasing for a coherent state input (as described in section (\ref{eqn:CS}) with $\abs{\alpha}=0.5$). Figure (b) shows the detection probability for a bare JPM and a JPM experiencing pure dephasing for a one photon Fock state input.}
\label{fig:T2Det}
\end{figure}

To further illuminate the effect of pure dephasing on a JPM we look at the probability of detection for a coherent state input and a one-photon Fock state input. In figure \ref{fig:T2Det}, we see that for both input states, dephasing suppresses the oscillations in detection probability exhibited by the bare JPM.  These oscillations result from coherent excitation swapping (Rabi-type oscillations) between the cavity and JPM, and superpositions of an excitation in the JPM and in the resonator are subject to dephasing processes in the JPM.  This pure dephasing turns coherent JPM-cavity oscillations into incoherent resonant tunneling.  

In the long time limit, both the dephased and bare JPM detect a photon with the same probability.   Thus, it is not necessary to aim at long $T_2$ values for a JPM as one would for a phase qubit \cite{Martinis02,Martinis09c}.  On the contrary, we see that the dephasing incurred by measurement is smaller at short $T_2$, rendering it advantageous. A more detailed discussion about $T_2$ as an engineering parameter will be given in the end of the paper.

\subsubsection{Energy Relaxation of the JPM}

We consider a JPM experiencing energy dissipation, as described by a master equation including the Lindblad operator $\hat{J}_3$ of equation (\ref{eqn:rel}). In this situation, the JPM will also experience associated dephasing on a timescale $T_2 = 2T_1$.  The $\chi^1$-matrix of a dissipating JPM has additional non-zero elements (in addition to the $\beta_{jk}$) attributed to a change of total excitation number.  
As the JPM can lose photons into an external heat bath, it is possible that multiple photons from the cavity might excite the JPM before a detection event occurs.   The nonzero $\chi^1$-matrix elements are of the form
\begin{equation}
\beta_{jk}^{(r)}\equiv\chi^1_{j-1j+rk-1k+r},
\end{equation}
corresponding to the loss of  integer $0<r<{\rm min}\{k,j\}$ photons before detection.  $\beta_{jk}^{(0)}=\beta_{jk}$ from the previous sections, and the diagonal elements are relabeled $\alpha^{(r)}_j\equiv\beta_{jj}^{(r)}$.

\begin{figure}[h]
\includegraphics[width = \columnwidth]{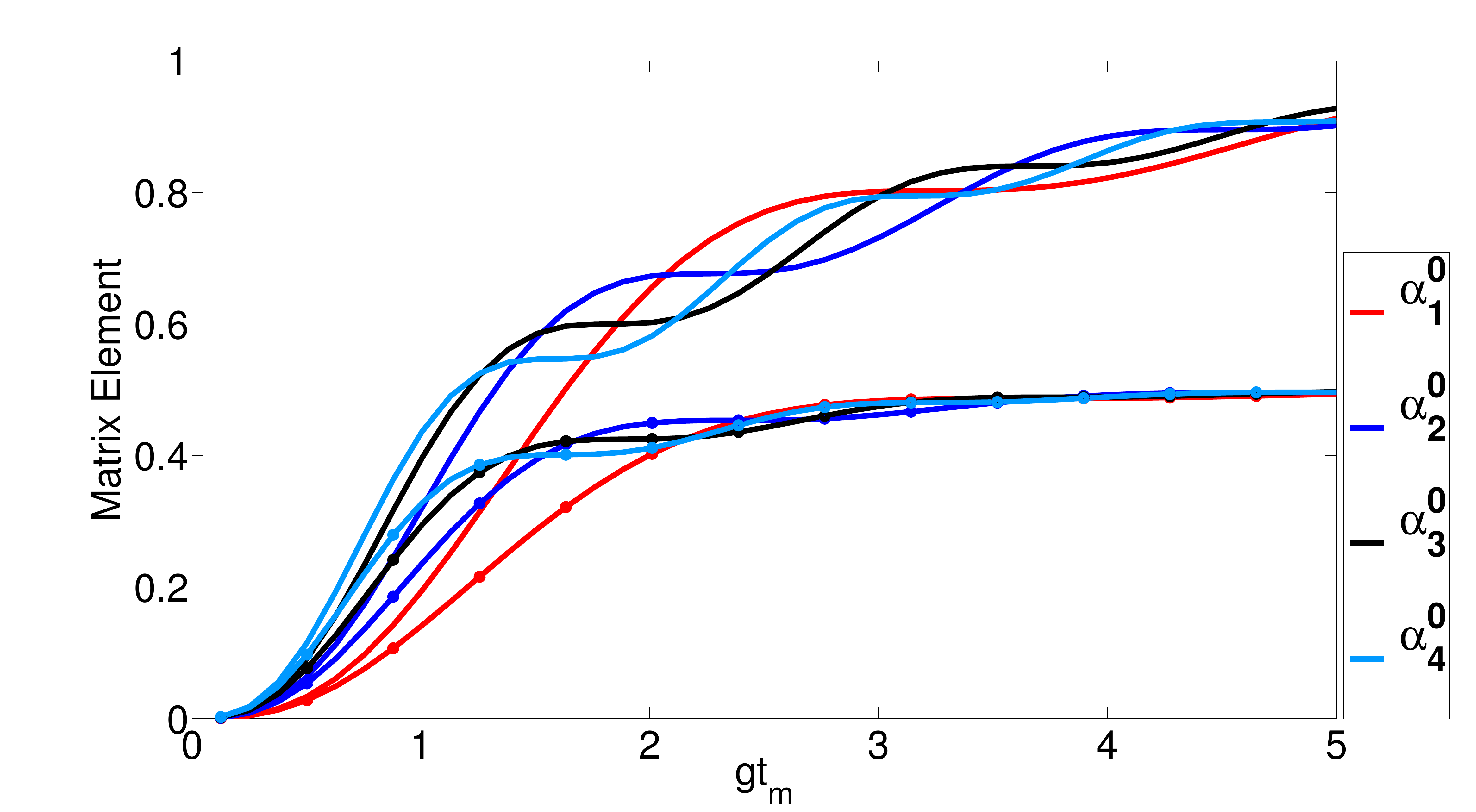}
\caption{This figure shows the diagonal $\chi^1$ matrix elements shared by a bare JPM and a JPM experiencing energy relaxation. Energy relaxation matrix elements are represented by cirlces on the plots.}
\label{fig:T1Diag}
\end{figure}

\begin{figure}
\includegraphics[width = \columnwidth]{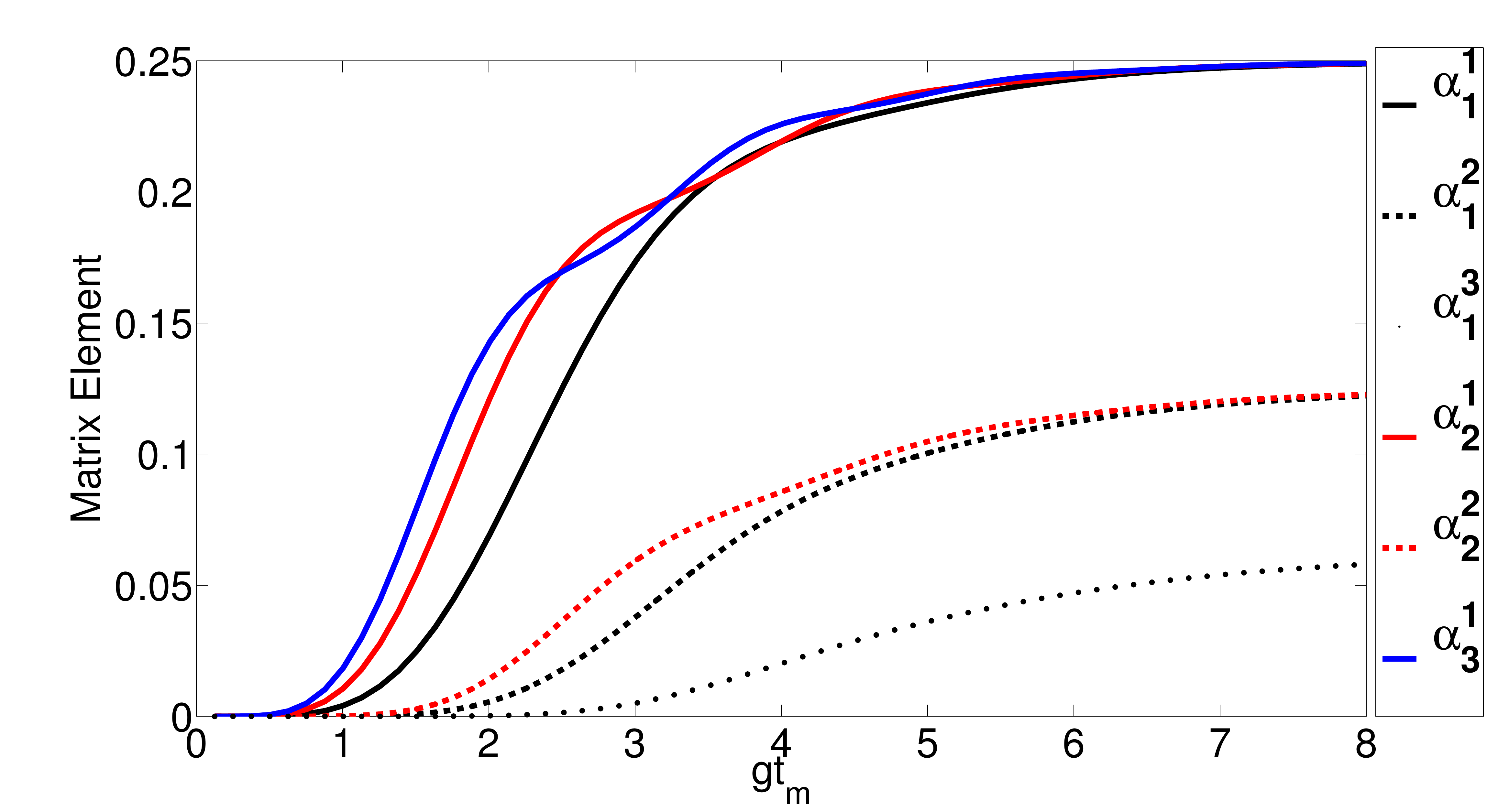}
\caption{Diagonal $\chi^1$-matrix elements $\alpha_j^{(r)}\equiv\chi^1_{j-1j+rj-1j+r}$ corresponding to photon detection after loss of $r$ photons.  These elements are zero unless an energy dissipation mechanism is present.}
\label{fig:T1UniDiag}
\end{figure}

In figure \ref{fig:T1Diag} we compare the  $\alpha_j^{(0)}$ for an energy relaxation timescale of $\frac{1}{T_1} = \gamma_1$ to the $\alpha_j$ of a detector with infinite $T_1$.  Unlike the case of pure dephasing, energy loss from the JPM reduces the asymptotic value of these $\chi^1$-matrix elements.
We also note that the off-diagonal $\beta_{jk}^{(0)}$ evolve exactly like those of a JPM experiencing pure dephasing, as shown in figure \ref{fig:T2Off}, but with an effective $T_2$ of  ${2T_1}$.
The diagonal $\chi^1$-matrix elements $\alpha_j^{(r)}$ are shown in figure \ref{fig:T1UniDiag} for different values of $r$.

\begin{figure}
\subfigure[Coherent State Detection Probability]{
\label{fig:T1CDet}
\includegraphics[width = \columnwidth]{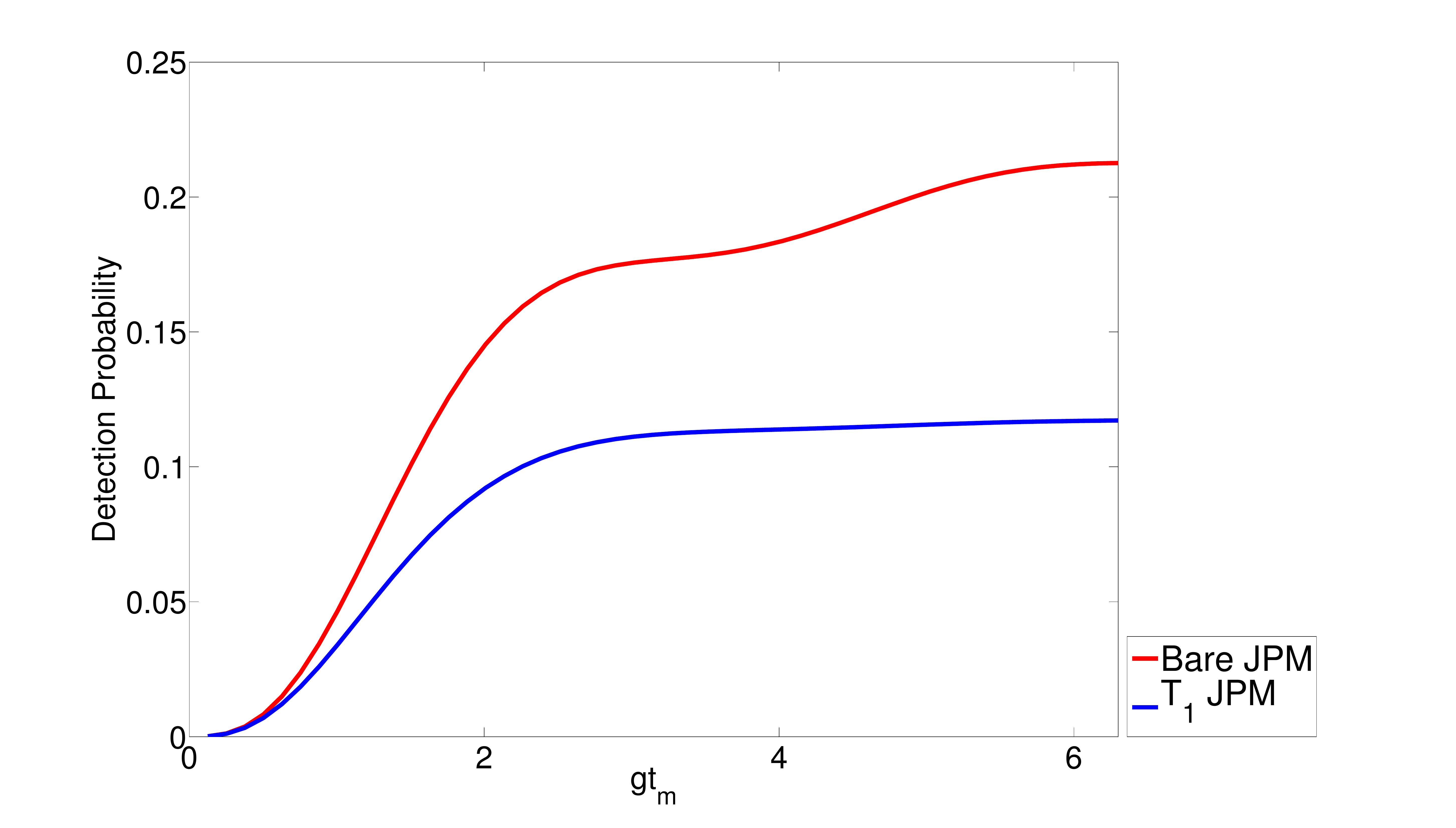}}
\subfigure[One Photon Fock State Detection Probability]{
\label{fig:T1FDet}
\includegraphics[width = \columnwidth]{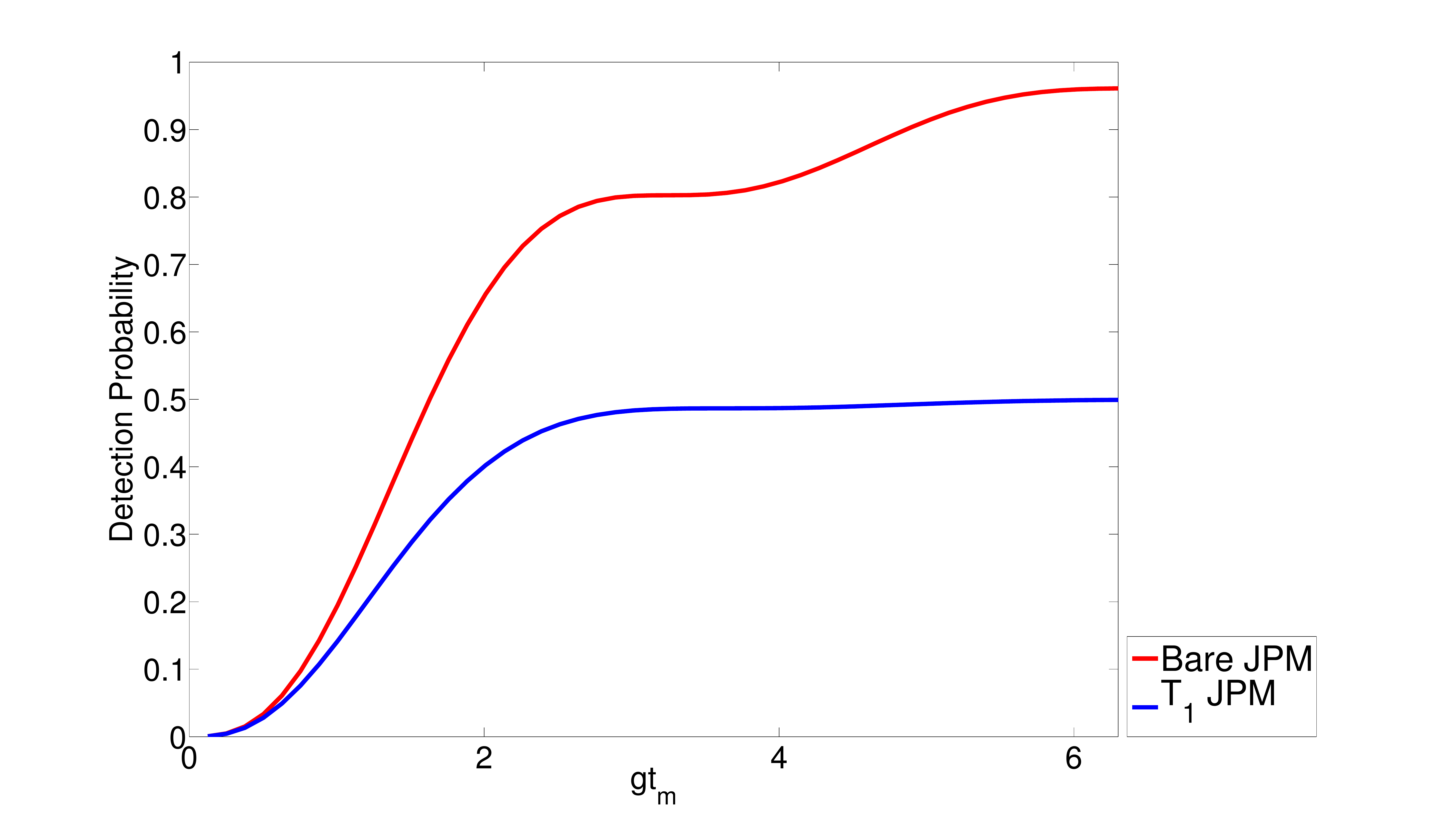}}
\caption{Figure (a) shows the detection probability for a bare JPM and a JPM experiencing energy relaxation for a coherent state input (same as in \ref{fig:T2CDet}). Figure (b) shows the detection probability for a bare JPM and a JPM experiencing energy relaxation for a one photon Fock state input (same as in \ref{fig:T2FDet}).}
\label{fig:T1Det}
\end{figure}

As in the case of pure dephasing, it is instructive to see the effect of energy relaxation on the probability of photon detection for specific input states. Figure \ref{fig:T1Det} shows the detection probability as a function of time for both coherent states and one photon Fock states.
The short time oscillatory behaviour of the bare JPM detection probability is strongly suppressed by energy relaxation because of the effective dephasing rate $T_2 = 2 T_1$. In addition to this dephasing effect, the detection probability of a JPM experiencing energy relaxation asymptotes more quickly to a smaller value than that of a bare JPM. 

\begin{figure}
\includegraphics[width = \columnwidth]{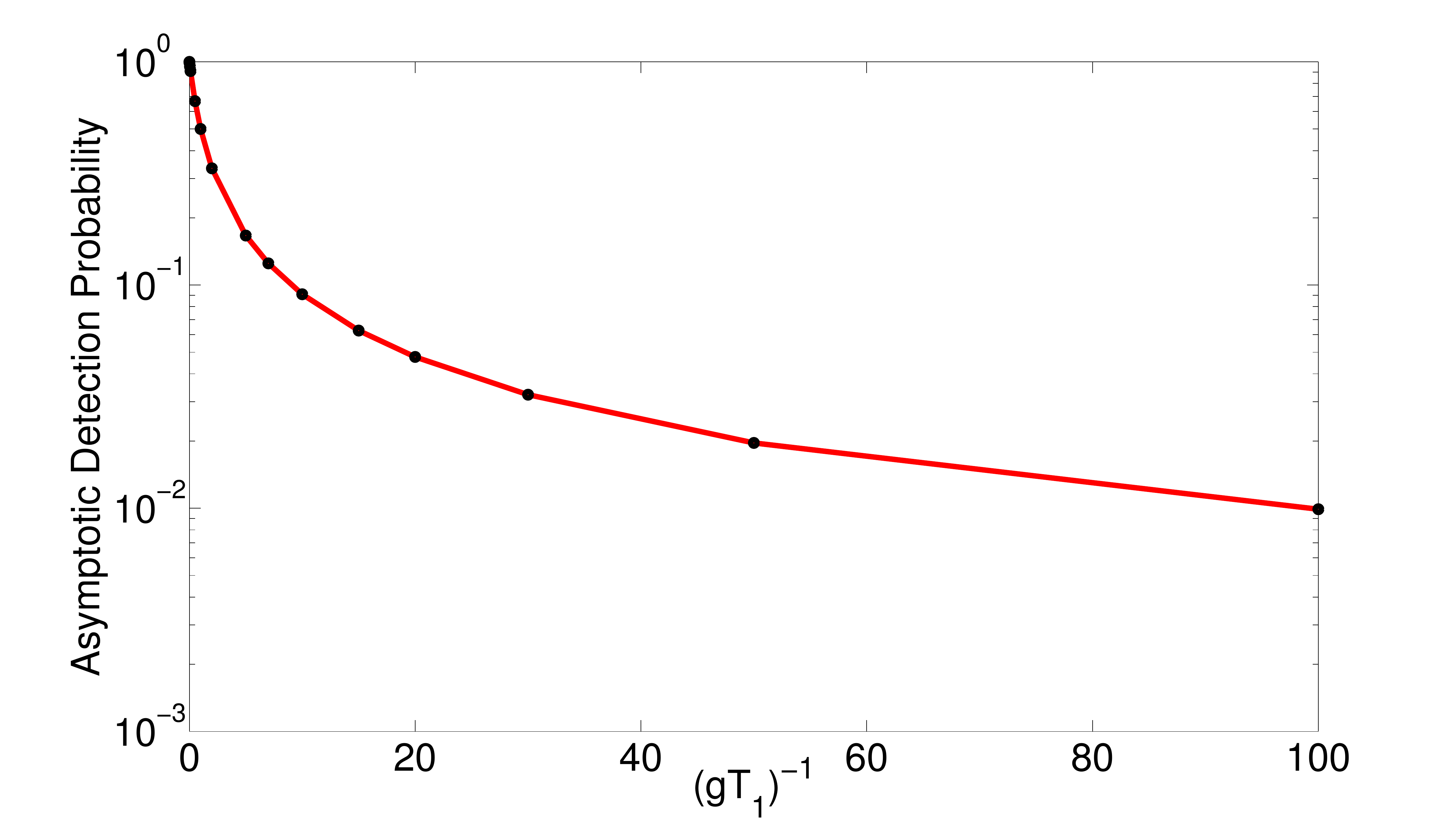}
\caption{This figure shows the asymptotic value of detection probability for a JPM experiencing energy relaxation as a function of the energy relaxation rate $T_1^{-1}$. The black circles represent numerically simulated data points, while the red curve is a linear fit between the simulated data points. As expected, it is a monotonically decreasing function.}
\label{fig:T1Asym}
\end{figure}

A plot of the asymptotic value of detection probability as a function of energy relaxation rate, $T_1^{-1}$, is shown in figure \ref{fig:T1Asym} for a one photon Fock state input.  The JPM experiences competing decay channels (energy relaxation and incoherent tunnelling into the measured state), only one of which results in photon detection.  This reduces the time required for the probability to reach its asymptotic value as well as the  overall probability that a detection will occur. In the case where these two decay rates are equal, the single photon detection probability will be exactly half of what it would be for a bare detector, as seen in figure \ref{fig:T1Det}.

\subsubsection{Dark Counts}

The final modification of interest is tunneling out of the $|0\big>_{\rm d}$ state, a detection event which does not change the number or excitations in the cavity and is therefore considered a dark count.  
This is described by a master equation including the Lindblad operator $\hat{J}_0$ of equation (\ref{eqn:DC}). 
 The $\chi^1$-matrix  has, in addition to  all the non-zero elements of the bare JPM, additional non-zero elements
\be
\label{eqn:DCel}
\chi^1_{jjkk} \neq 0 \; \forall\;  j,k\in\{0,1,..N-1\}.
\ee
These elements correspond to detection events that occur without changing the $k$-photon Fock state in the cavity.
The JPM experiencing dark counts will appear to have a higher probability of photon detection than a bare JPM; however, this increased probability is due to false detections. 
Dark counts limit the detector contrast by the ratio between true and false detections.

We are not aware of any simple way to correct for all the effects of dark counts on detection probability without a priori information about the detected state; however, by the appropriate choice of experimental parameters, the dark count rate can be made to be quite small for a JPM -- as much as 2 to 3 orders of magnitude smaller than the excited state tunneling rate \cite{Chen:2011ys}. It is therefore not unreasonable to simulate a dark count rate of 5 \% of the excited state tunneling rate as a conservative estimate. 

\begin{figure}
\subfigure[One Photon Fock State Detection Probability]{
\label{fig:DCFDet}
\includegraphics[width = \columnwidth]{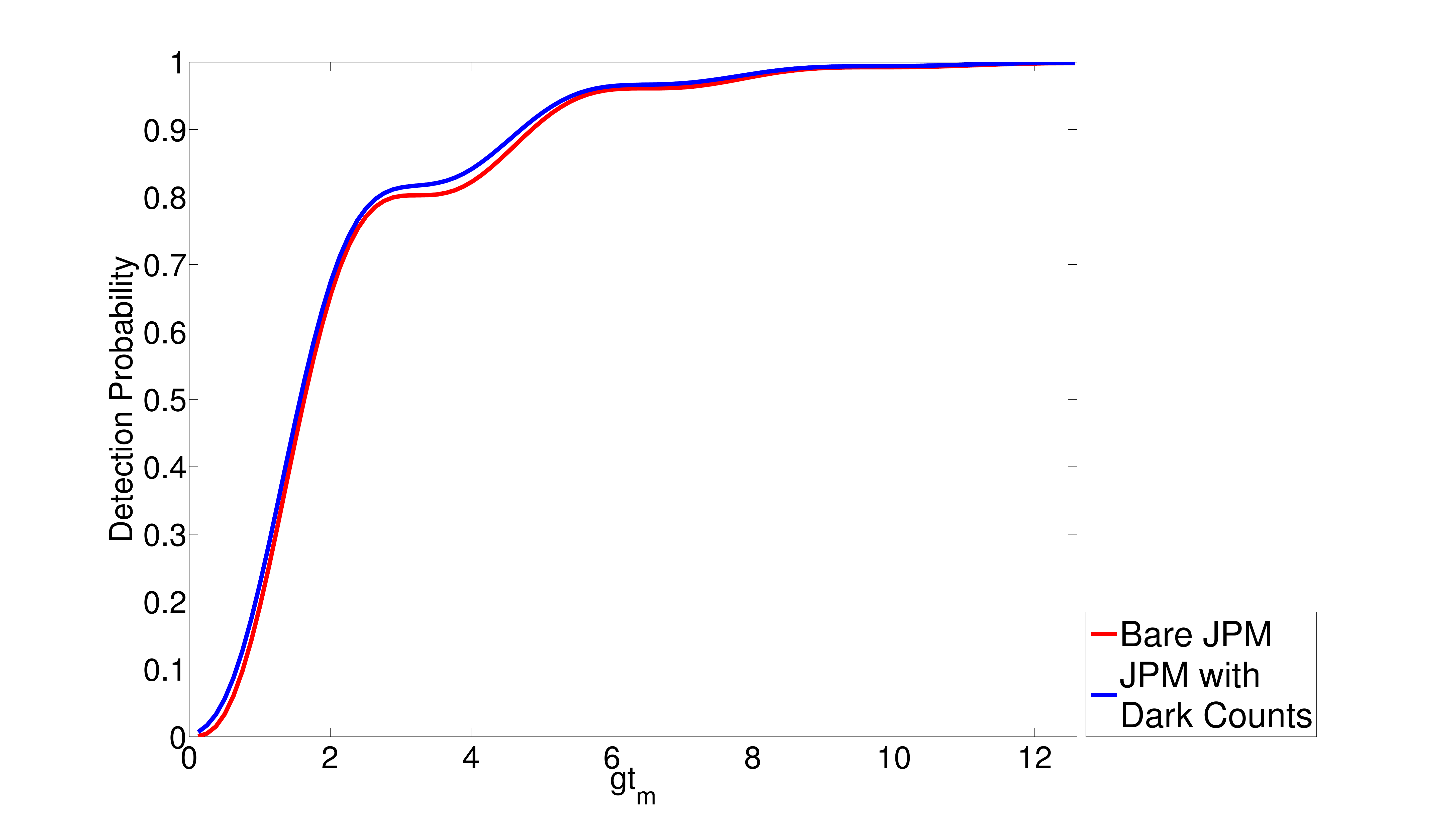}}
\subfigure[Coherent State Detection Probability]{
\label{fig:DCCDet}
\includegraphics[width = \columnwidth]{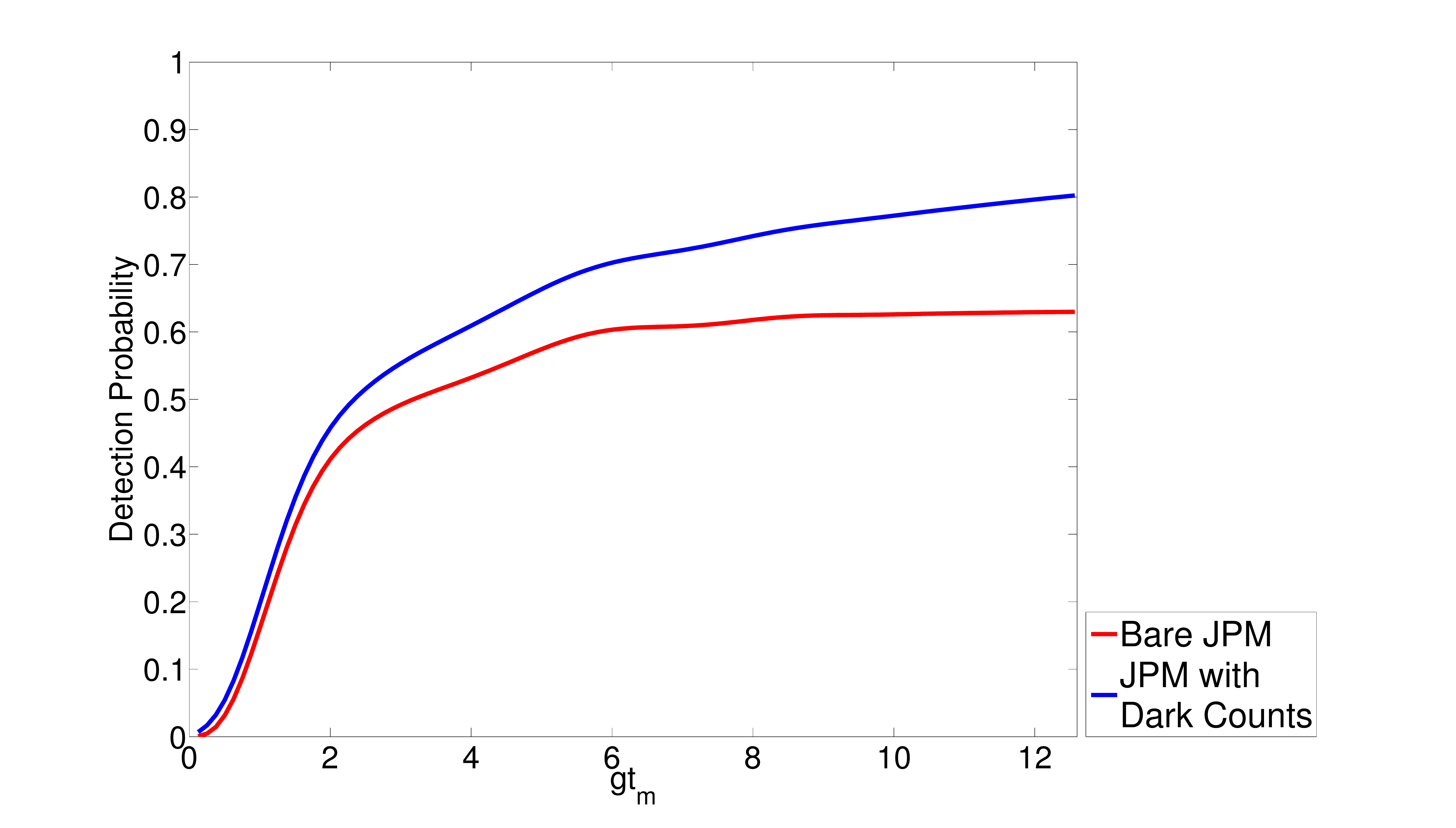}}
\caption{Detection probabilities for a bare JPM and one experiencing dark counts for (a) a one-photon Fock state input state and (b) an $\alpha=1$ coherent state input state.}
\label{fig:DCDet}
\end{figure}

Figure \ref{fig:DCDet} shows the detection probability for a one-photon Fock state input with a 5\% dark count rate, which is very similar to that of a bare JPM; however, for a coherent state with $\big<n\big>=|\alpha|^2=1$, dark counts significantly change the detection probability since the coherent state has a significant vacuum component. This deviation decreases as $|\alpha|$ of the coherent state increases.

\subsection{Analytical Solutions in the Low ${\rm T_2}$ Regime}
\label{sec:Analytics}
While a full analytic solution of the system's master equation does not promise more illumination than the numerical results presented above, here we obtain the short $T_2$ behavior of the detector by making appropriate approximations.
We begin by defining the block cavity matrix 
\be
\rho_{ij}(t) \equiv \bra{i}_{\rm d} \xi(t) \ket{j}_{\rm d}
\ee
 where $\ket{i}_{\rm d}\equiv \mathbb{I}_{\rm c}\otimes\ket{i}_{\rm d}$ project $\xi(t)$ on the basis states of the noninteracting JPM.   
We are interested in the unrenormalized state of the cavity after photon detection, which corresponds to the cavity state $\rho_{mm}(t)$. From the full master equation,  it can be shown that
\be
\label{eqn:RedMeas}
\dot{\rho}_{mm}(t)  = \bra{m}_{\rm d} e^{{S}t}\xi(0) \ket{m}_{\rm d} = \gamma_1\rho_{11}(t)
\ee
in the case that $\gamma_0=0$. Thus, the instantaneous time evolution of $\rho_{mm}$ depends only on $\rho_{11}$, the unnormalized state conditioned on the JPM being in the excited state.

The time evolution of $\rho_{11}$ is governed by a system of four first-order operator differential equations:
\bea
\dot{\rho}_{11}&=&ig\left(\rho_{10}\hat{a}^\dagger-\hat{a} \rho_{01}\right)-\gamma_1 \rho_{11}\\
\dot{\rho}_{00}&=&ig\left(\rho_{01} \hat{a} - \hat{a}^\dagger \rho_{10}\right)\nonumber\\
\dot{\rho}_{01}&=&ig\left(\rho_{00}\hat{a}^\dagger-\hat{a}^\dagger \rho_{11}\right)-\frac{\gamma_1 + \kk_2}{2}\rho_{01}\nonumber\\
\dot{\rho}_{10}&=& ig\left(\rho_{11}\hat{a}-\hat{a}\rho_{00}\right)-\frac{\gamma_1 + \kk_2}{2}\rho_{10}\nonumber
\label{eqn:Red4}
\eea
\noindent where we have introduced the pure dephasing rate $\kk_2 = \frac{1}{T_2}$ for notational convenience. 
This system can be reduced to a single fourth order operator differential equation in terms of $\rho_{11}(t)$,  as shown in equation (\ref{eqn:Red1}).

Now for simplicity, we consider the occupation probability of the cavity state $\ket{n}\bra{n}_{\rm c}$ given that a photon detection has occurred,
\bea
\label{eqn:RedMeasN}
P_{n}(t) &\equiv& \int_0^t\bra{n}_{\rm c}\dot{\rho}_{mm}(t')\ket{n}_{\rm c}dt' ,
\eea
which from (\ref{eqn:RedMeas}) can be expressed as 
\bea
\label{eqn:RedMeasN}
&&P_{n}(t)  = \gamma_1\int_0^t x(t') dt',
\eea
where we have defined the matrix element
\bea
\label{eqn:RedMeasN}
 x(t)\equiv  \bra{n}_{\rm c}\rho_{11}(t)\ket{n}_{\rm c}.
\eea
The full master equation reduces to the fourth order differential equation in $x(t)$, given in (\ref{eqn:RedOrd}).

Before attempting to solve the somewhat cumbersome equation (\ref{eqn:RedOrd}), we make the simplifying assumption that $T_2$ is the smallest time scale of the system's evolution, and keep only the highest order terms in ${\frac{g}{\kk_2},\frac{\gamma_1}{\kk_2}\ll1}$. This gives 
\bea
\label{eqn:RedApprox}
x^{(4)} &+& \kk_2 x^{(3)} + \frac{1}{4}\kk_2^2x^{(2)}+ \nonumber\\
 &\; & \frac{1}{4}\yy_1\kk_2^2x^{(1)} + \kk_2 g^2 \yy_1 (1+n)x = 0,
\eea
which can be solved with Laplace transforms.  Defining $X(s) \equiv \mathcal{L}[x(t)]$, equation (\ref{eqn:RedApprox}) in the Laplace domain is 
\bea
\label{eqn:Lap}
X(s) &=& 2g^2(n+1)x_0 \left(s + \frac{\kk_2 - 3\gamma_1}{2}\right)\\
&\times&\left(s^4 + \kk_2 s^3 + \frac{1}{4}(\kk_2^2 s^2 +\kk_2^2 \yy_1 s) + \kk_2 g^2 \yy_1 (n + 1)\right)^{-1}\nonumber
\eea
where $x_0 \equiv x(0) = \bra{n}_{\rm c}\rho_{11}(0)\ket{n}_{\rm c}$.

While still very general and valid for all input states, equation (\ref{eqn:Lap}) is still somewhat unwieldy.  
However, in the limit of short $T_2$, we can assume that coherent oscillations between the cavity and JPM become incoherent tunneling.  Defining $\Gamma_{\alpha,\beta\rightarrow\delta,\gamma}$ to be the tunneling rate from state $|\alpha\big>_{\rm c}\otimes|\beta\big>_{\rm d}$ to state $|\gamma\big>_{\rm c}|\otimes|\delta\big>_{\rm d}$, we take
\bea
\nonumber&&\Gamma_{{\rm n},0 \rightarrow {\rm n} - 1,1}  = \Gamma_{{\rm n}-1,1 \rightarrow {\rm n},0}  = 4ng^2T_2 \\
&&\Gamma_{\rm n,1 \rightarrow n,m}  = \yy_1
\label{eqn:FermiRates}
\eea
Here $\Gamma_{\rm n,0 \rightarrow n - 1,1}$ is the incoherent tunnelling rate from the cavity into the JPM when $n$ photons are present,  and $\Gamma_{\rm n-1,1 \rightarrow n,0}$ the rate for the inverse process. Both rates are broadened by short $T_2$, which can be understood in terms of the Purcell effect.   
If we consider only  Fock state inputs, the occupation probabilities 
\be
P_{n,{\rm j}}(t) \equiv \bra{n}_{\rm c}\rho_{\rm jj}(t)\ket{n}_{\rm c} \ \ {\rm j} \in \{0,1,{\rm m}\}
\ee
of the cavity being in the $n$-photon Fock state and the detector being in state $\ket{{\rm j}}_{\rm d}$ obey the Pauli master equation.  Using the rate in  (\ref{eqn:FermiRates}),  this simplifies to 
\bea
\nonumber\dot{P}_{n,0} &=& n\yy_2\left(P_{n-1,1} - P_{n,0}\right) \\
\nonumber\dot{P}_{n,1} &=& (n+1)\yy_2\left(P_{n+1,0} - P_{n,1}\right) - \yy_1 P_{n,1} \\
\dot{P}_{n,{\rm m}} &=& \yy_1 P_{n,1}
\label{eqn:FermiEqns}
\eea
where $\yy_2 \equiv (2g)^2T_2 \equiv {(2g)^2}/{\kk_2}$. For an $n$-photon Fock state input, $P_{n,0}(0)=1$ and the total number of excitations in the system is fixed to $n$, so at later times only 
 $P_{n,0}$, $P_{n-1,1}$ and $P_{n-1,{\rm m}}$ are nonzero. 
 
 The equations (\ref{eqn:FermiEqns}) can be solved to find the detection probability as a function of time (the details of this are shown in Appendix \ref{app:LowT2}), giving
\bea
P_{n,{\rm m}}(t) &=& 1+  \frac{\yy_1 \yy_2n}{(s_+-s_-)}\left( \frac{e^{s_+t}}{s_+} - \frac{e^{s_-t}}{s_-}\right)
\label{eqn:LowT2Det}
\eea
where
\bea
s_{\pm} = \frac{1}{2}\left( -\yy_1 - 2\yy_2n \pm \sqrt{\yy_1^2 + 4\yy_2^2n^2} \right).
\label{eqn:LowT2roots} 
\eea
We can distinguish two regimes for this solution. In the tunneling-limited regime, $\gamma_1\ll \gamma_2 n$ we find $s_{\rm +}\simeq -\gamma_1/2$ and $s_{\rm -}\simeq -2\gamma_2n$ and have
\be 
P_{n,m}(t)=1-e^{-\gamma_1 t/2}+O(\gamma_1/\gamma_2n)
\ee
In the opposite regime, photon capture is the slower, limiting process ($\gamma_1\gg \gamma_2 n$). In this regime, $s_+ = -\yy_2n$ and $s_- = -\yy_1 - \yy_2n$ and we find
\be
P_{n,m}=1-e^{-\gamma_2nt}+O(\gamma_2n/\gamma_1)
\ee
The differences between these regimes is evident in figure 3(a);  with added dephasing, $\gamma_1\gg\gamma_2 n$, and the detection probabilities $P_{n,m}=\alpha_n$ reach their asymptotic values more slowly and with $n$-dependence.  Without the added dephasing, $\gamma_1\ll\gamma_2 n$, the rate at which the detection probabilities reach their asymptotic value is determined by $\gamma_1$, and independent of $n$. 

\begin{figure}
\subfigure[Low $T_2$ Regime]{
\label{fig:LowT2DetProb}
\includegraphics[width = \columnwidth]{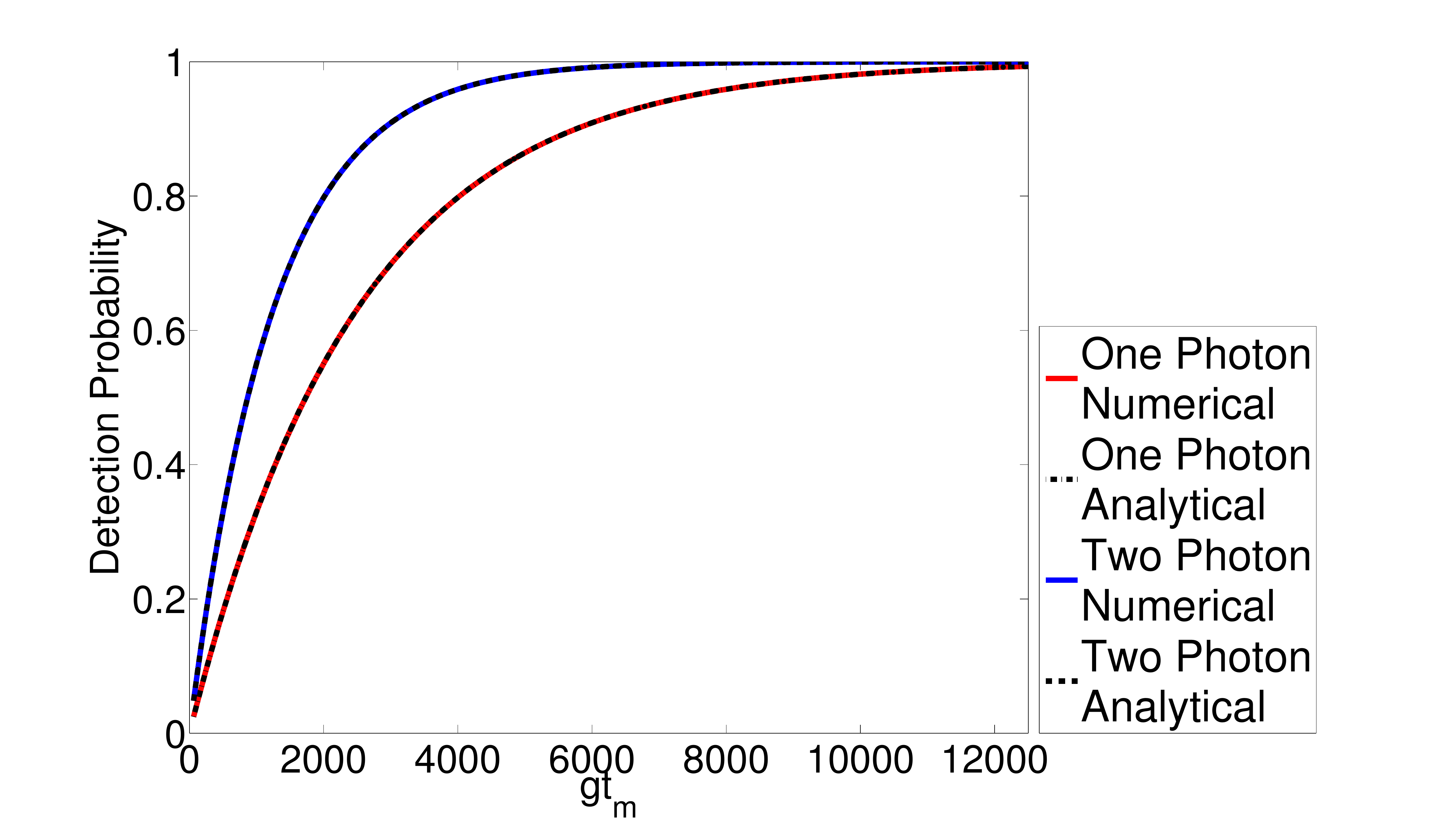}}
\subfigure[Intermediate $T_2$ Regime]{
\label{fig:IntT2DetProb}
\includegraphics[width = \columnwidth]{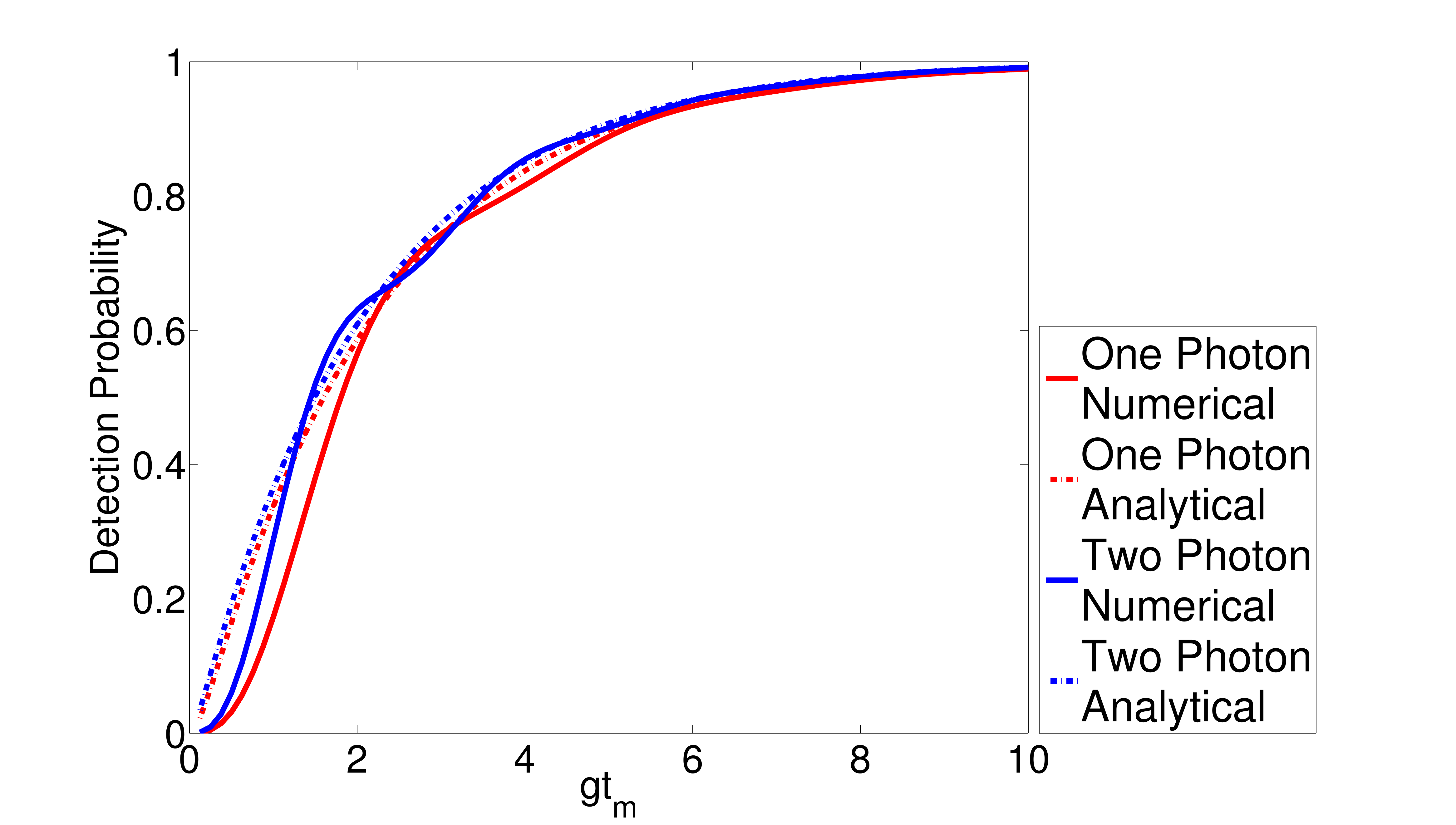}}
\caption{These figures show the detection probability of a JPM experiencing pure dephasing in the low and intermediate $T_2$ regimes for one and two photon Fock state inputs. The detection probability obtained from the analytical solution described in this section is compared to a numerical simulation (via the Liouville supermatrix approach) of the detection probability. In figure (a) $\frac{1}{T_2} = 10000\gamma_1$ and in figure (b) $\frac{1}{T_2} = \gamma_1$.}
\end{figure}

Equation (\ref{eqn:LowT2Det}) agrees to a high degree of accuracy with the numerical simulations for a JPM experiencing pure dephasing with a very short dephasing time, $T_2$. As can be seen in figure \ref{fig:LowT2DetProb} for one and two photon Fock states, this agreement occurs at all times.
Interestingly, even at longer $T_2$ the analytical solution is still a good approximation to the numerical simulation. By design, the analytical solution ignores coherent oscillations between the cavity and the JPM, and so in the long $T_2$ regime the analytical solution does not display the oscillatory behaviour of the numerical solution, but instead describes its average behavior. This can be seen in figure \ref{fig:IntT2DetProb}, for one and two photon Fock state inputs.
 
\section{Coherent State Test}

In this section, we propose a test to determine whether a given JPM's back action is closer to the lowering operator or the subtraction operator in order to correctly characterize the JPM.  Additionally, such a test would allow us to examine the effects of energy relaxation and pure dephasing on the time evolution of the back action.

\subsection{The Dependence of Detection Probability on Coherent State Power}

Consider a coherent state expressed in the Fock basis,
$
\label{eqn:CS}
\ket{\alpha} = e^{-\frac{\abs{\alpha}^2}{2}} \sum_{n = 0}^{\infty} \frac{\alpha^n}{\sqrt{n!}} \ket{n}.
$
It is straightforward to calculate the detection probabilities for both the lowering and subtraction operator back actions, as a function of $\alpha$:
\bea
&&\nonumber{\rm Lowering \  Operator} \\
\label{eqn:LowScale}
&& P_{{\rm low}}(\alpha) \equiv {\rm Tr}\left[ \hat{a}\ket{\alpha} \bra{\alpha}\hat{a}^\dagger \right] =\abs{\alpha}^2 \\ 
\nonumber\\
&&\nonumber{\rm Subtraction \ Operator} \\
\label{eqn:SubScale}
&&\nonumber P_{{\rm sub}}(\alpha) \equiv {\rm Tr}\left[ \hat{B}_{\rm m}\ket{\alpha} \bra{\alpha}\hat{B}_{\rm m}^\dagger \right] \\
&&= 1 - e^{-\abs{\alpha}^2}=|\alpha|^2-\frac{|\alpha|^4}{2}+O\left(|\alpha|^6\right)
\eea
As equations (\ref{eqn:LowScale}) and (\ref{eqn:SubScale}) show, the difference in the detection probabilities occurs only in the nonlinear response, i.e., terms of the order of $|\alpha|^4$. This is consistent with the observation that $\hat{a}$ and $\hat {B}$ have identical matrix elements up to $1$ photon but are different at higher photon numbers. By measuring the detection probability for coherent states of varying power and examining how this detection probability scales with the power of the coherent state, it is possible to characterize the back action with respect to the lowering operator and the subtraction operator.

\subsection{Bare JPM Coherent State Test}

First we apply this test to our simulations of a bare JPM, recalling that we expect $\hat{B}^1(t) \sim \hat{a}$ at short times (see equation (\ref{eqn:B1Order})) and $\hat{B}^1(t) \sim \hat{B}_{\rm m}$ at long times (see discussion in sect. \ref{sec:notunnel}).  The proportionality constants affect the detection probability and not the structure of the back action, so we remove them by renormalizing (\ref{eqn:LowScale}) and (\ref{eqn:SubScale}) as follows:
\bea
\tilde{P}_{{\rm low}}(\alpha) =  \frac{P_{{\rm low}}(\alpha) P_{{\rm data}}(\alpha = \alpha_0)}{P_{{\rm low}}(\alpha = \alpha_0)} \\
\tilde{P}_{{\rm sub}}(\alpha) =  \frac{P_{{\rm sub}}(\alpha) P_{{\rm data}}(\alpha = \alpha_0)}{P_{{\rm sub}}(\alpha = \alpha_0)} 
\eea 
where $P_{{\rm data}}$ is the simulated detection probability, and $\alpha_0$ is the smallest value of $\alpha$ with simulated data.  The $\alpha$-scaling of the JPM can be more directly compared to those of these rescaled lowering and subtraction operators. Experimentally, this is important as it accounts for calibration uncertainties that may occur, such as from attenuation or imperfect impedance matching.

\begin{figure*}
\subfigure[$gt_m = 0.126$ Detection Probability]{
\label{fig:IdScale01}
\includegraphics[width = \columnwidth]{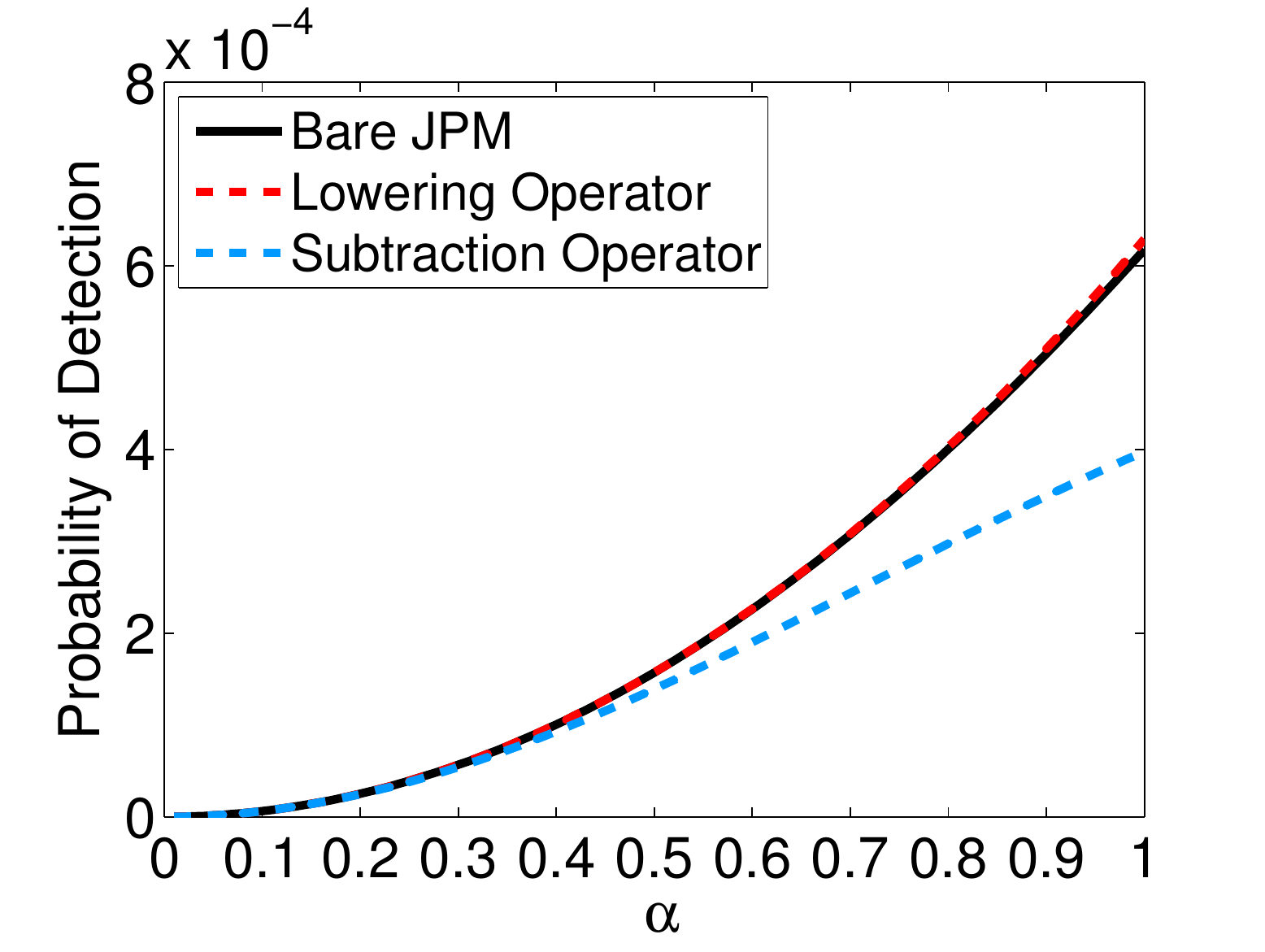}}
\subfigure[$gt_m = 1.26$ Detection Probability]{
\label{fig:IdScale1}
\includegraphics[width = \columnwidth]{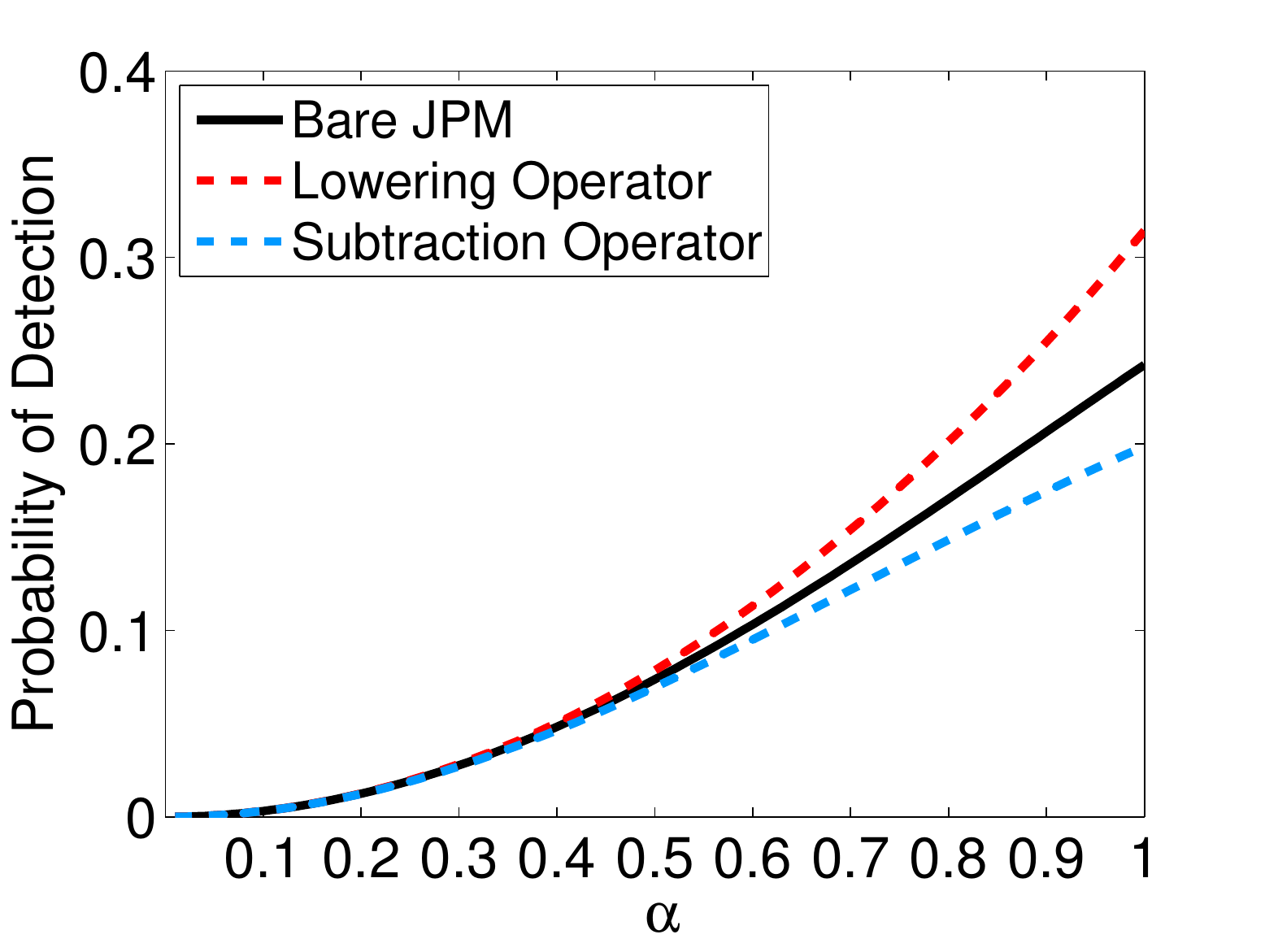}}
\subfigure[$gt_m = 2.52$ Detection Probability]{
\label{fig:IdScale2}
\includegraphics[width = \columnwidth]{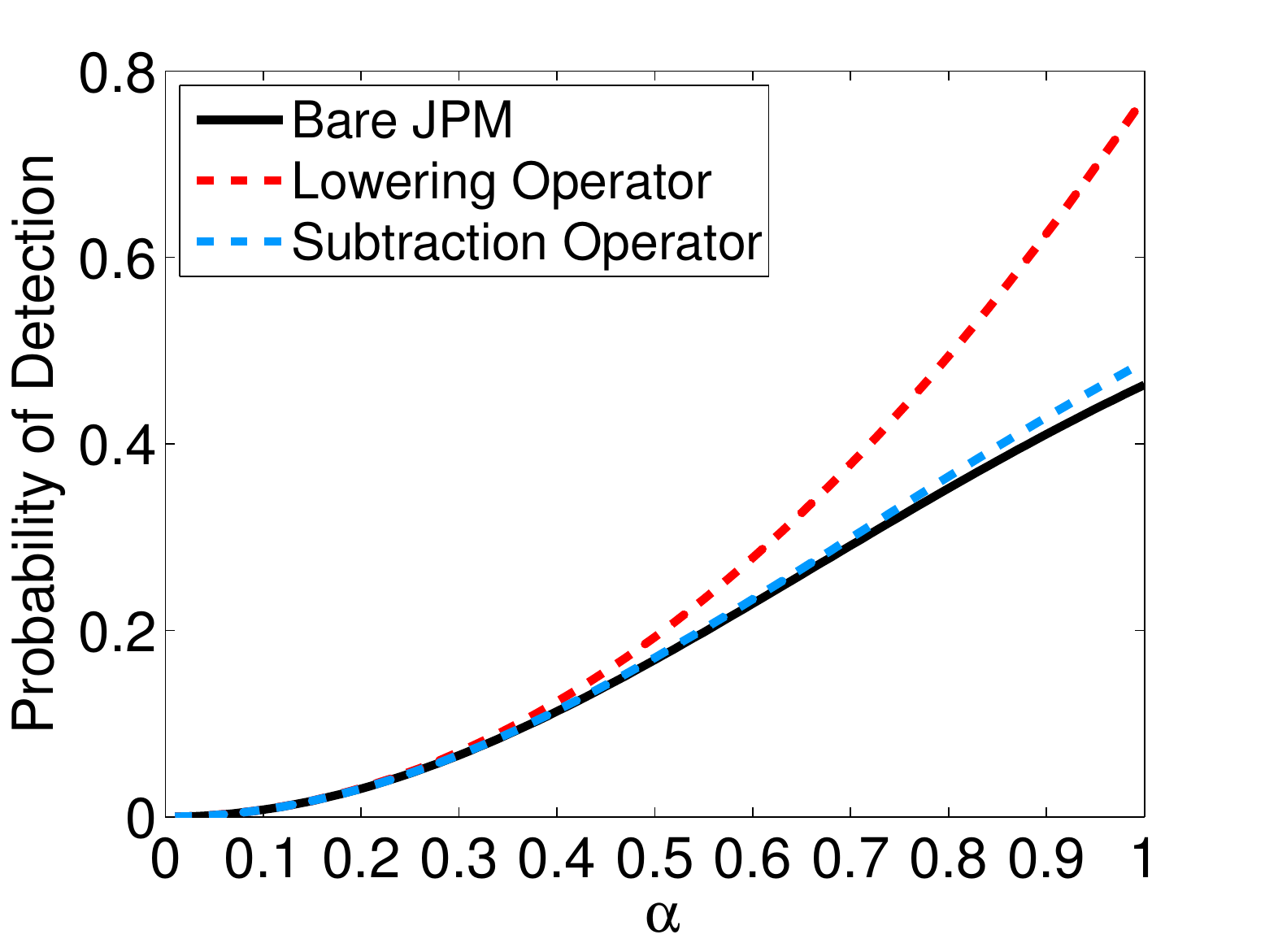}}
\subfigure[$gt_m = 12.6$ Detection Probability]{
\label{fig:IdScale10}
\includegraphics[width = \columnwidth]{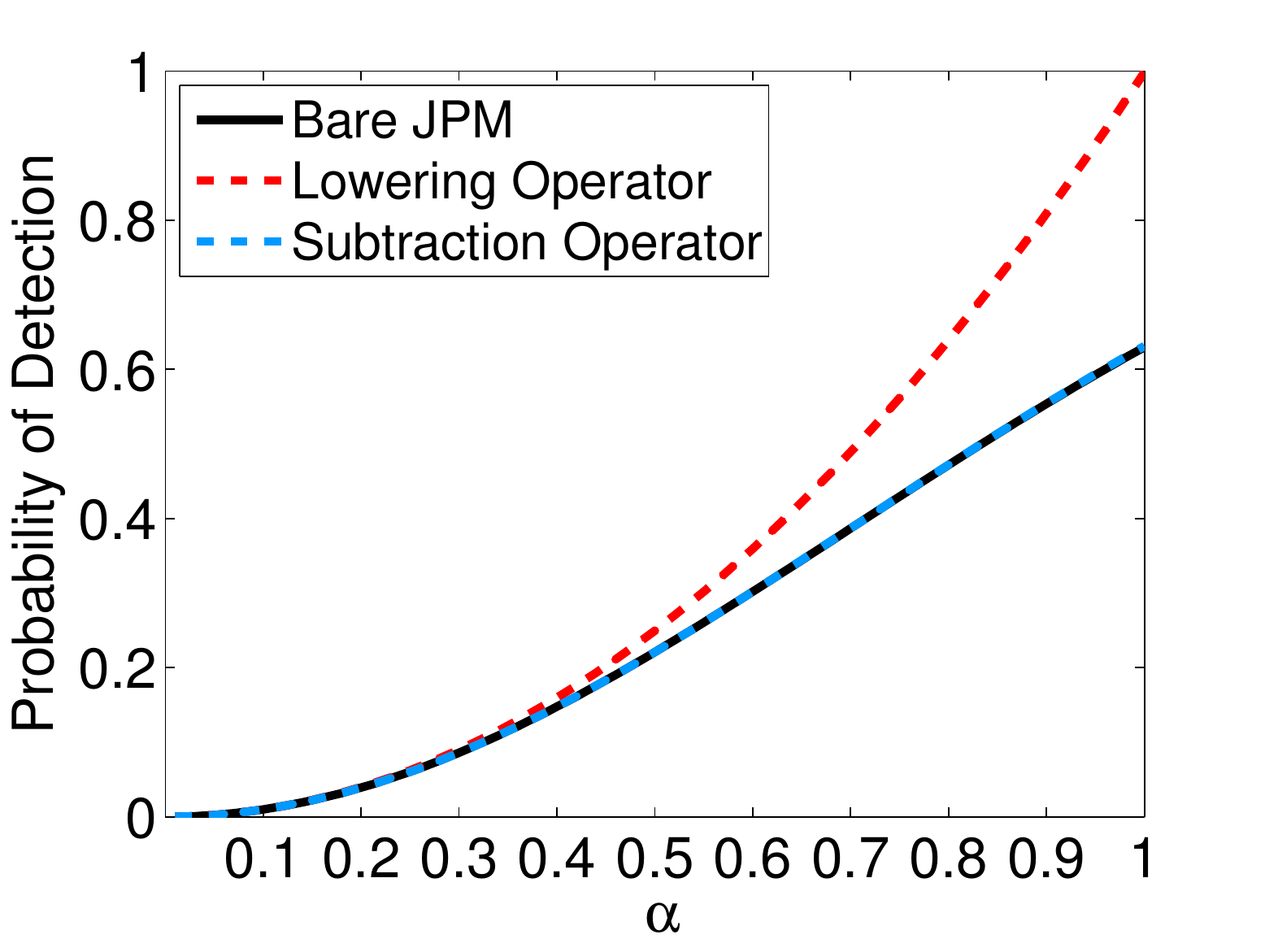}}
\caption{For a bare JPM, the detection probability is shown for a measurement occurring at $t_mg$ values of $0.126$ in (a), $1.26$ in (b), $2.52$ in (c) and $12.6$ in (d). $\alpha$ runs from $\alpha = 0.01$ to $\alpha =1$ in 0.01 intervals, and in each figure, all three curves are scaled to be equal at $\alpha = 0.01$. The time in (d) is chosen such that it is representative of the long time steady state of the back action.}
\label{fig:IdScale}
\end{figure*}

Figures \ref{fig:IdScale01} and \ref{fig:IdScale10} show the detection probabilities of a bare JPM as a function of the power of the coherent input state $\alpha$ for a $gt_m$ value of $0.126$ and $12.6$ respectively.   As expected, the back action is very close to the lowering operator at short times -- only deviating slightly at high powers --  and at long times the back action is very close to the subtraction operator.  At short times, the deviation from the lowering operator at high powers can be explained by equation (\ref{eqn:B1Order}), which only predicts the back action will be proportional to $\hat{a}$ for $\sqrt{n}gt \ll 1$. Thus, at high powers (large $|\alpha|=\sqrt{n}$), corrections to (\ref{eqn:B1Order}) will become important, as is this case for measurement times long compared to  $t_{\rm crit}=(g\sqrt{n})^{-1}$.
Figures \ref{fig:IdScale1} and  \ref{fig:IdScale2} show the back action at intermediate times.  While \ref{fig:IdScale1} shows behavior intermediate between the lowering and subtraction operators, 
figure \ref{fig:IdScale2} shows that $\alpha$-scaling can actually fall below that of the subtraction operator. This effect is a result of the additional dephasing incurred in measurement. 

\subsection{The Effects of Pure Dephasing and Energy Relaxation}

\begin{figure*}
\subfigure[$gt_m = 0.126$ Detection Probability]{
\label{fig:T1Scale01}
\includegraphics[width = \columnwidth]{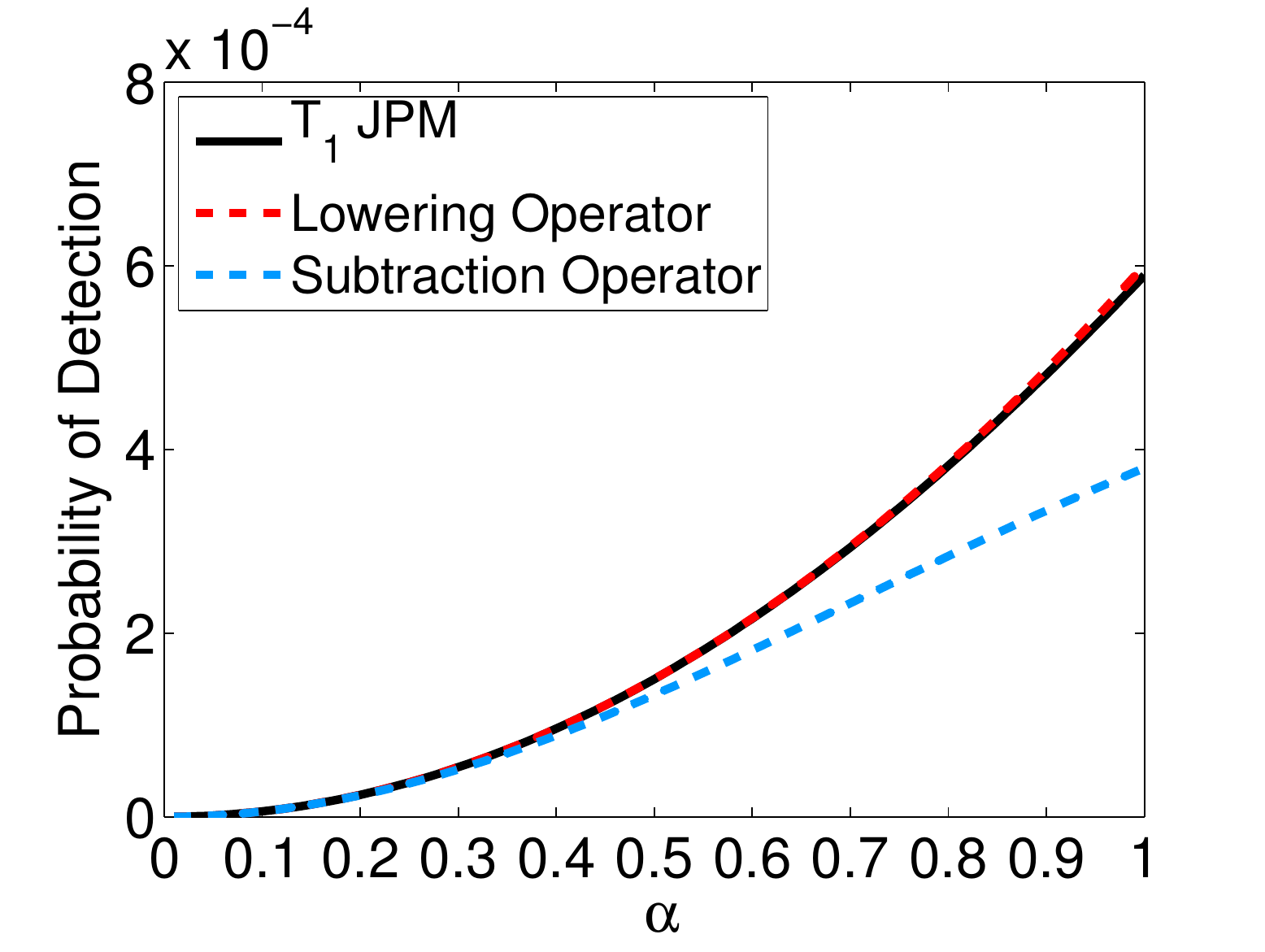}}
\subfigure[$gt_m = 1.26$ Detection Probability]{
\label{fig:T1Scale1}
\includegraphics[width = \columnwidth]{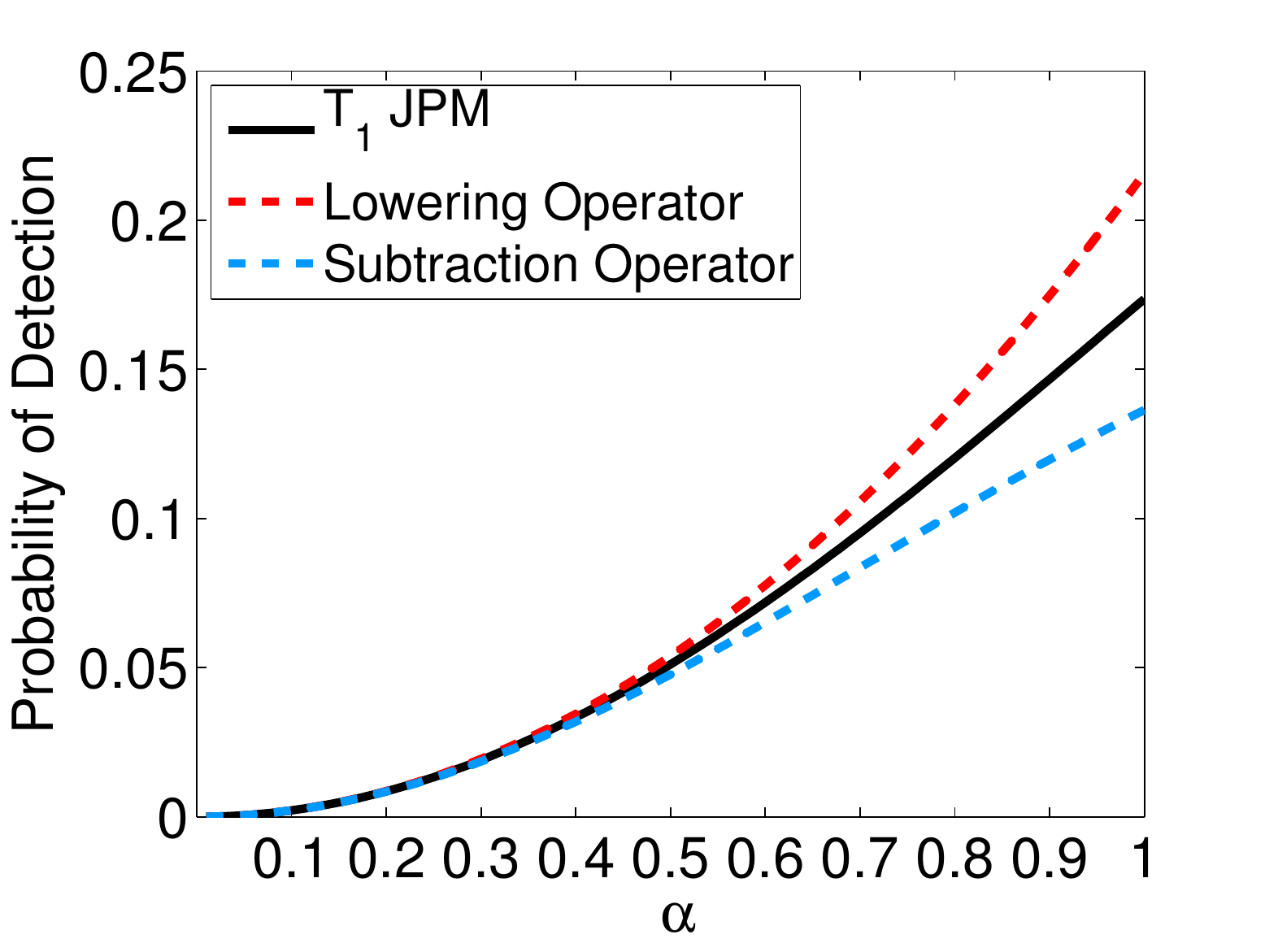}}
\subfigure[$gt_m = 2.52$ Detection Probability]{
\label{fig:T1Scale2}
\includegraphics[width = \columnwidth]{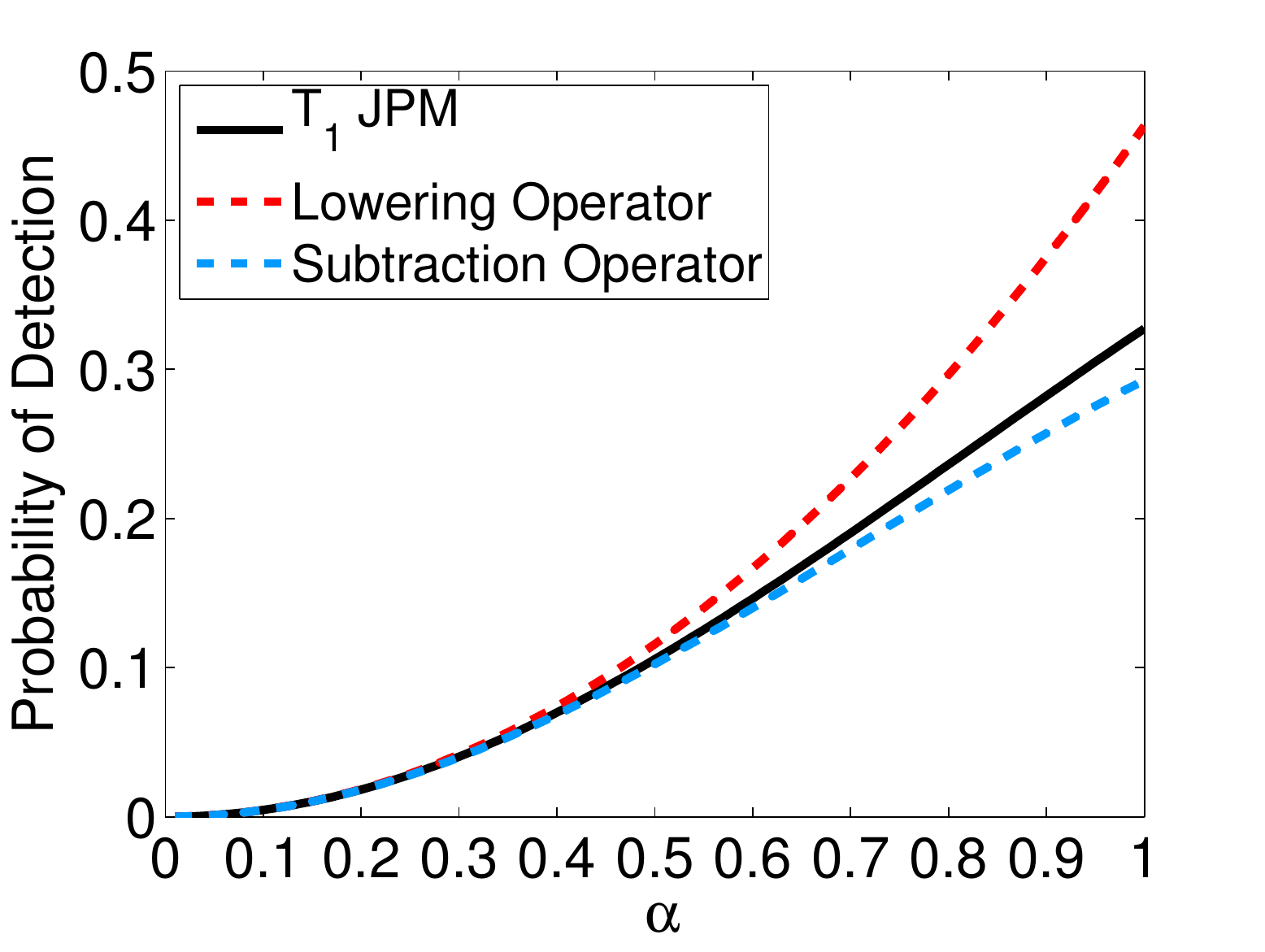}}
\subfigure[$gt_m = 12.6$ Detection Probability]{
\label{fig:T1Scale10}
\includegraphics[width = \columnwidth]{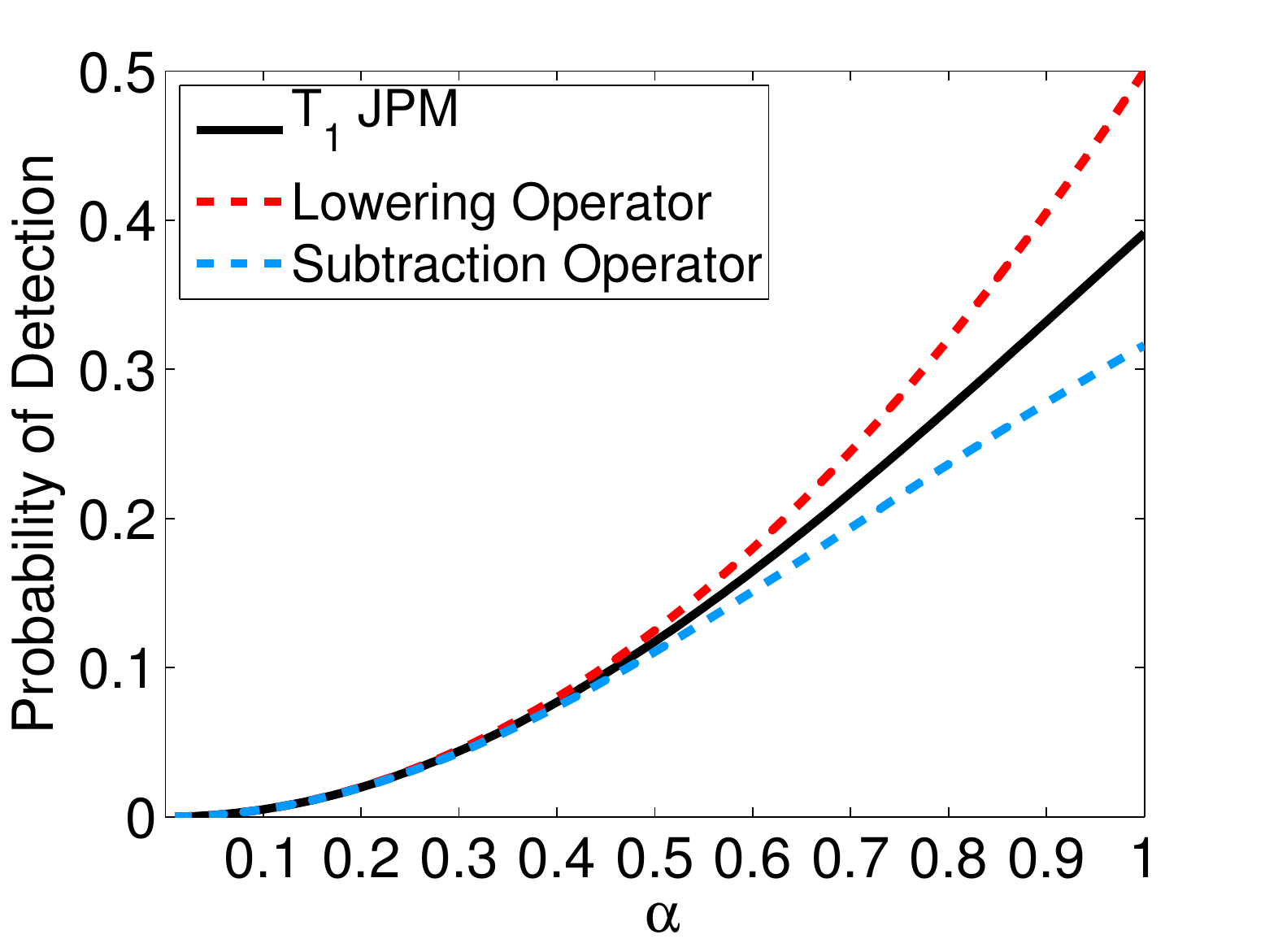}}
\caption{For a JPM experiencing energy relaxation (with $T_1 = \frac{1}{\yy_1}$ as before), the detection probability is shown for a measurement occurring at $t_mg$ values of $0.126$ in (a), $1.26$ in (b), $2.52$ in (c) and $12.6$ in (d). $\alpha$ runs from $\alpha = 0.01$ to $\alpha =1$ in 0.01 intervals, and in each figure, all three curves are scaled to be equal at $\alpha = 0.01$.}
\label{fig:T1Scale}
\end{figure*}

We now study the affects of energy relaxation and pure dephasing on the $\alpha$-scaling of the coherent state test. Figure \ref{fig:T1Scale} shows the detection probability of a JPM experiencing energy relaxation at a rate $\frac{1}{T_1} = \gamma_1$ as a function of $\alpha$, at the same times as those of figure \ref{fig:IdScale}.  As can be seen in figure \ref{fig:T1Scale10} (which represents the long time steady state of the back action), the major effect of energy relaxation is to prevent the back action from fully transitioning to the subtraction operator at long times, but rather it asymptotes to an operator in the  intermediate regime.   With the additional energy relaxation channel present, the JPM detection probability becomes more sensitive to the number of photons in the cavity and does not fully approach the subtraction operator, which cannot resolve photon number.

\begin{figure*}
\subfigure[$gt_m = 0.126$ Detection Probability]{
\label{fig:T2Scale01}
\includegraphics[width = \columnwidth]{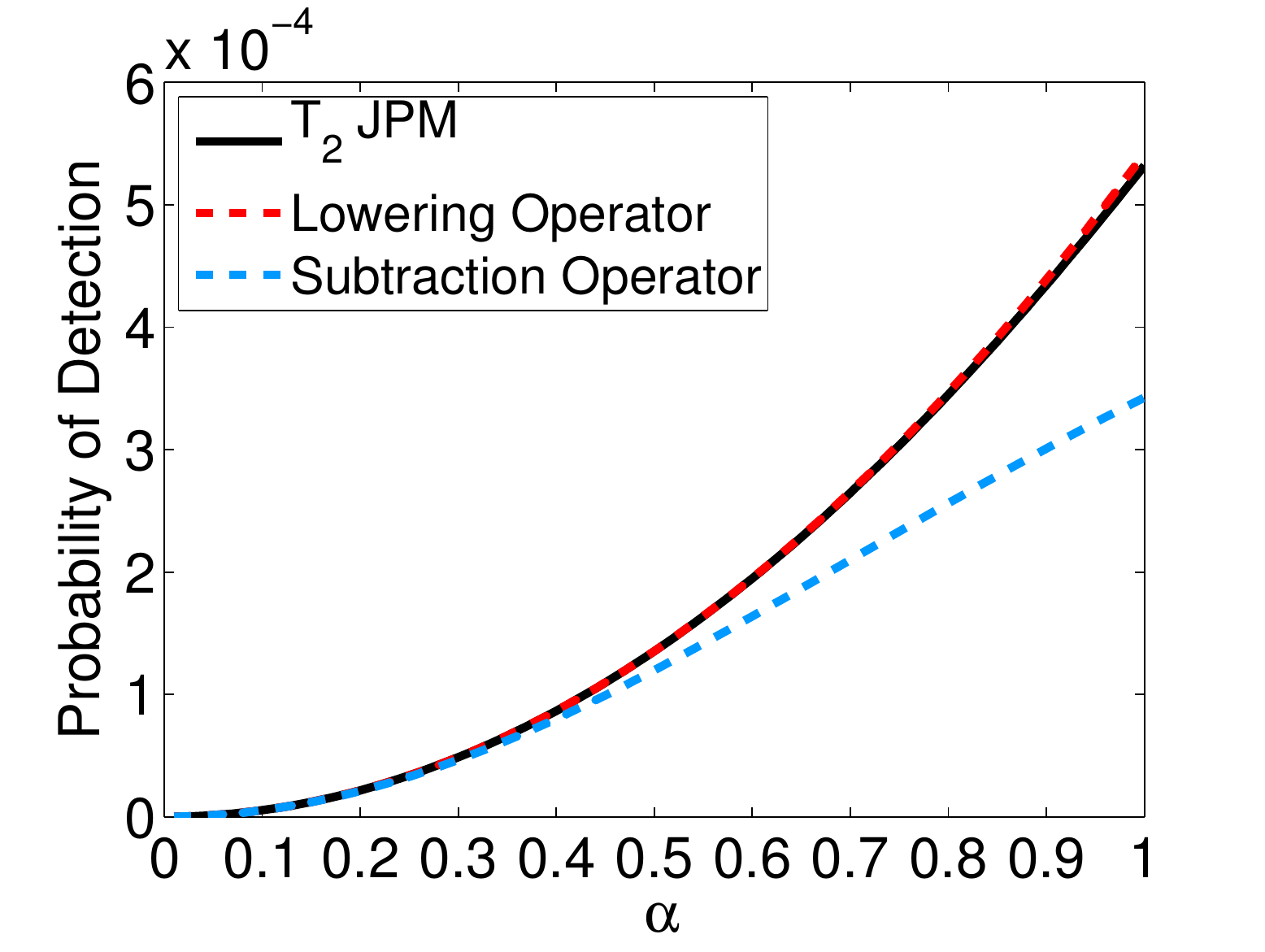}}
\subfigure[$gt_m = 1.26$ Detection Probability]{
\label{fig:T2Scale1}
\includegraphics[width = \columnwidth]{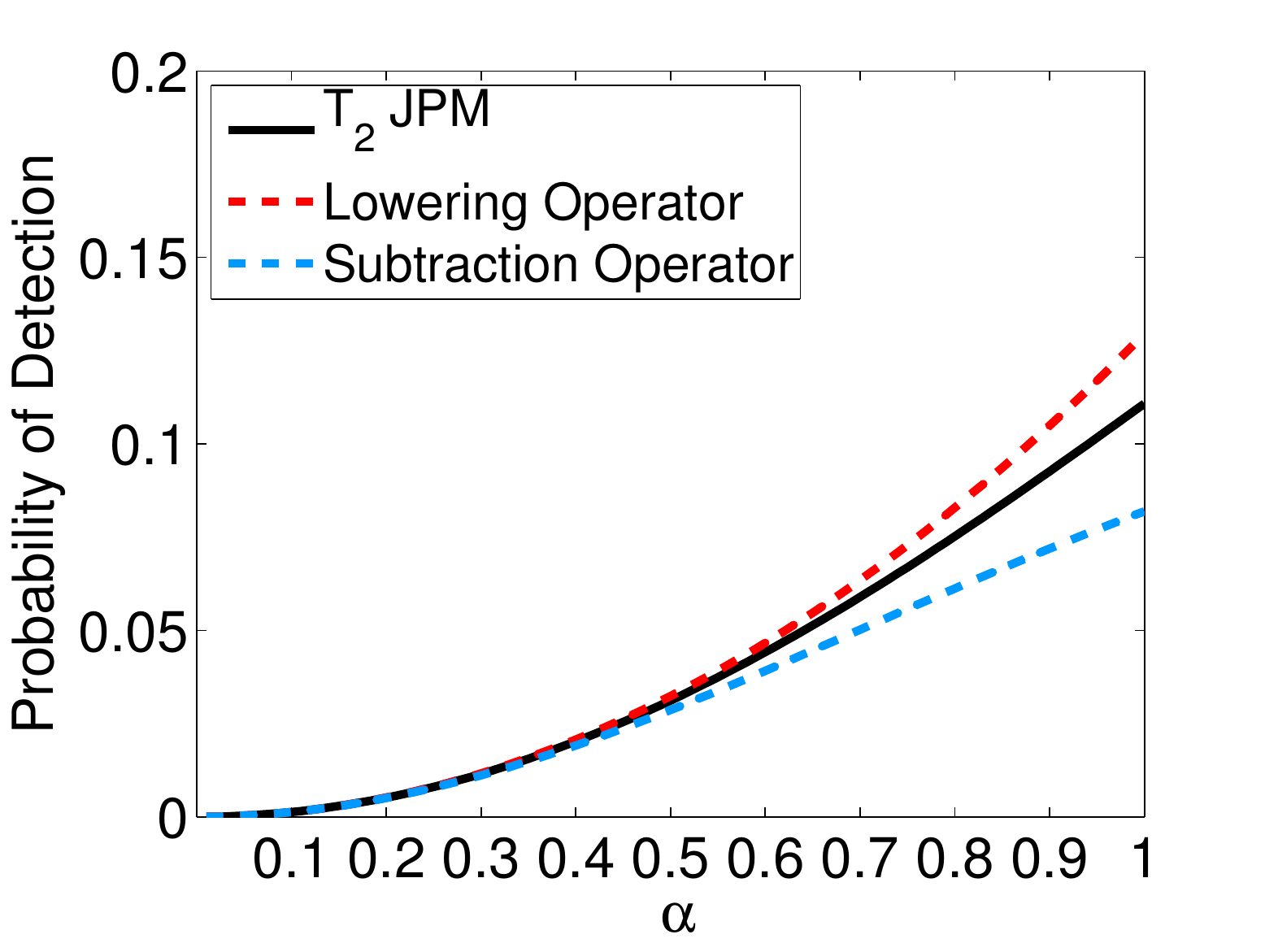}}
\subfigure[$gt_m = 2.52$ Detection Probability]{
\label{fig:T2Scale2}
\includegraphics[width = \columnwidth]{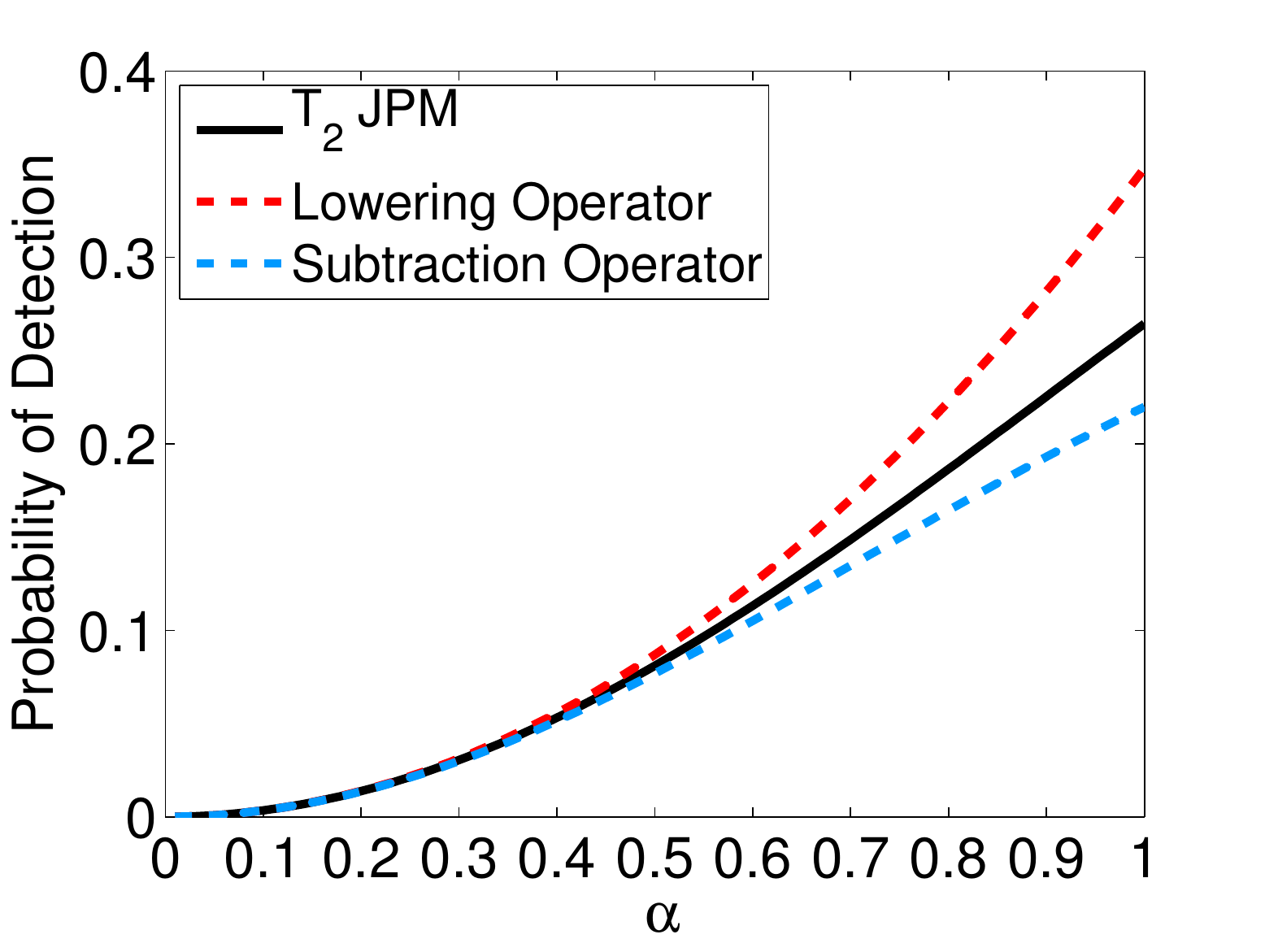}}
\subfigure[$gt_m = 12.6$ Detection Probability]{
\label{fig:T2Scale10}
\includegraphics[width = \columnwidth]{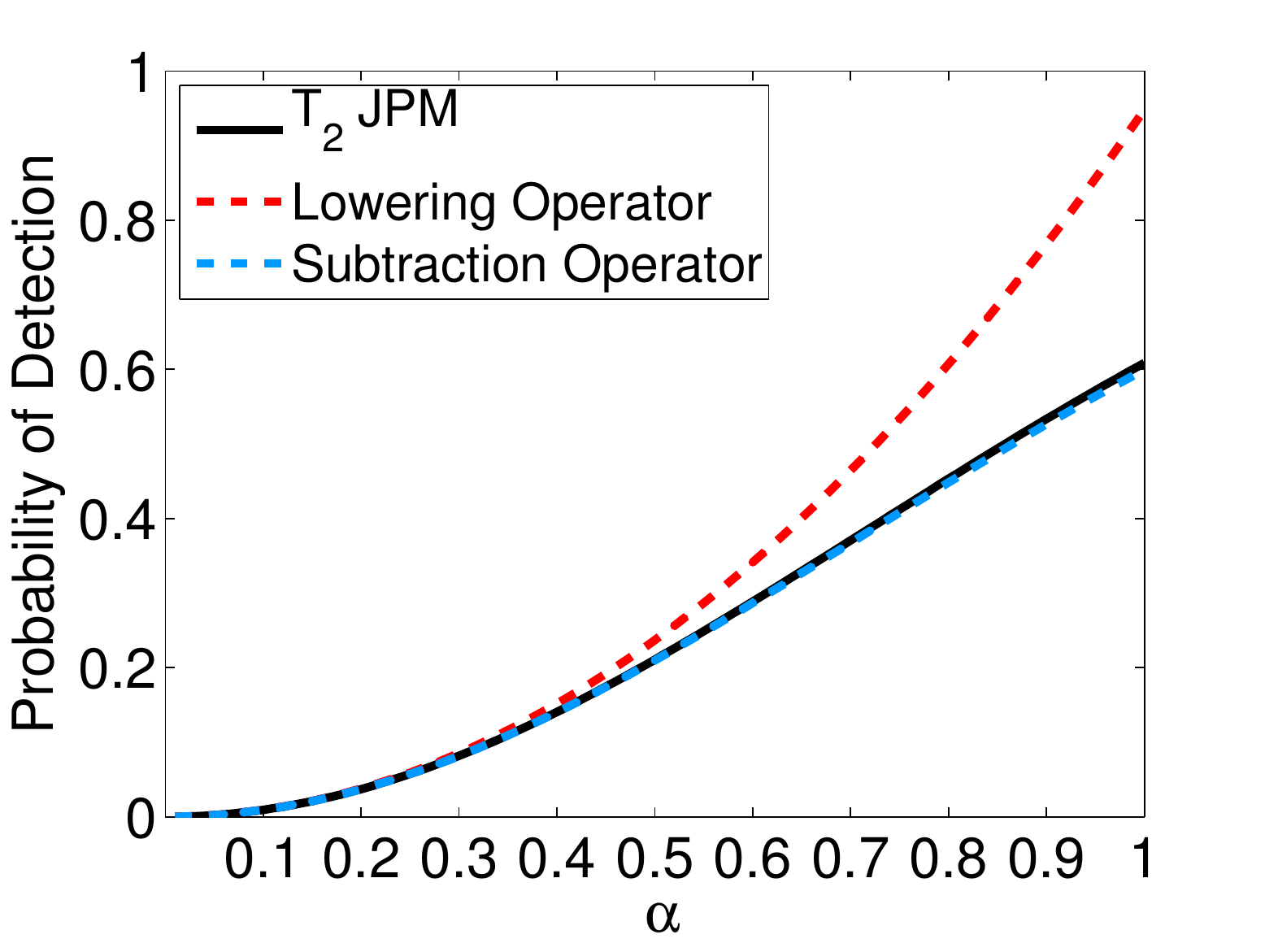}}
\caption{For a JPM experiencing pure dephasing (with $T_2 = \frac{10}{\yy_1}$ as before), the detection probability is shown for a measurement occurring at $t_mg$ values of $0.126$ in (a), $1.26$ in (b), $2.52$ in (c) and $12.6$ in (d). $\alpha$ runs from $\alpha = 0.01$ to $\alpha =1$ in 0.01 intervals, and in each figure, all three curves are scaled to be equal at $\alpha = 0.01$.}
\label{fig:T2Scale}
\end{figure*}

In addition, energy relaxation suppresses the sub-subtraction scaling at intermediate times (see figure \ref{fig:T1Scale2}).  We expect this is due to the added dephasing at $T_2=2T_1$, since 
figure \ref{fig:T2Scale} shows  that dephasing alone suppresses the drop below that of the subtraction operator at intermediate times (figure \ref{fig:T2Scale2}).
As can be seen, the effect of pure dephasing on $\alpha$-scaling is similar to that of energy relaxation; however, instead of stopping the back action from transitioning to the subtraction operator, pure dephasing merely increases the time scale on which this transition occurs.

\section{Optimal Regime for a JPM}

Albeit based on a similar circuit, the bare operation conditions for the JPM are different from those of a phase qubit, where long $T_1$ and $T_2$ are highly desirable. Operating a JPM at extremely long $T_2$ leads to the phenomenon of oscillating detection probability (figure \ref{fig:Bare}), sub-subtraction back action (figure \ref{fig:IdScale2}) and additional dephasing of the cavity (figure \ref{fig:BareOff}). This additional dephasing is undesirable as it destroys coherences in the original cavity state, irreversibly reducing its off-diagonal matrix elements, hence limiting the information available in a repeated measurement. 

It is thus advisable to operate the JPM in the short $T_2$ regime.  However, the effective photon-detector transfer rate, eq. (\ref{eqn:LowT2Det}), should be much shorter than the decoherence rate of the cavity, and this places a lower bound on $T_2$.   On the other hand, $T_1$-processes always limit the measurement fidelity and should be avoided. One way to achieve the limit of long $T_1$ with short $T_2$  is to damp the JPM with a frequency-dependent impedance with lowpass character, e.g., along the lines of \cite{Robertson05}, by shunting the JPM with an $LR$-element.

\section{Conclusions}

In this paper we have analyzed the back action of a JPM on the microwave cavity state it measures.  Numerical investigations of the cavity $\chi$ matrix conditioned on a detection event give us a convenient quantitative description of the detection process while including several relevant environmental processes.  At short times, the back action of a bare JPM is similar to the lowering operator, while at long times, its back action approaches the subtraction operator with additional cavity dephasing.   This additional dephasing can be reduced by adding pure dephasing to the JPM, which dampens the coherent oscillations between the JPM and the cavity without compromising the purity of the cavity state.  Energy relaxation decreases the asymptotic value of the diagonal cavity $\chi$ matrix elements and the detection probability by a factor of ${\yy_1}/({\yy_1 + T_1^{-1}})$.
It is useful to develop a test to determine which regime the JPM is operating in for different measurement times, and  the coherent state test is one such test that is straightforward to implement.  

\section*{Acknowledgements}
This work has greatly benefited from a wealth of discussions with Robert F. McDermott and Yung-Fu Chen as well as from early discussions with Jay Gambetta. 
We acknolwedge financial support of DARPA through the QuEST program. STM acknowledges support of IARPA through the MQCO program, LCGG was supported by NSERC and OGS grants and DP by the NSERC USRA program.

\appendix
\section{Low $T_2$ Analytics}
\label{app:LowT2}

The four first order operator differential equations of equation (\ref{eqn:Red4}) can be reduced to the following fourth order operator differential equation:
\begin{widetext}
\bea
\nonumber \rho_{11}^{(4)} &+& (\kk_2 + 2\yy_1)\rho_{11}^{(3)} + \Bigg[ \left( \frac{\kk_2 + \yy_1}{2}\right) \left( \frac{\kk_2 + 5\yy_1}{2}\right) +  2 {\cal D}_0 
+ 4g^2 \Bigg]\rho_{11}^{(2)} + \left[ (\kk_2 + 2\yy_1)(2g^2 + {\cal D}_0) + \yy_1\left( \frac{\kk_2 + \yy_1}{2} \right)^2\right]\rho_{11}^{(1)} \\ 
&+& \Bigg[ 4g^2 + {\cal D}_0\circ{\cal D}_0 + 2g^2\yy_1\left( \frac{\kk_2 + \yy_1}{2}\right) + 4g^2{\cal D}_0 + \yy_1\left( \frac{\kk_2 + \yy_1}{2}\right){\cal D}_0 \Bigg]\rho_{11} - 4g^4\hat{a}\hat{a}^\dag\rho_{11}\hat{a}\hat{a}^\dag = 0.
\label{eqn:Red1}
\eea
\end{widetext}
\noindent The superoperator ${\cal D}_0 [f] \equiv g^2\{\hat{a}^\dag \hat{a},f \} = g^2(\hat{a}^\dag \hat{a}f + f\hat{a}^\dag \hat{a})$ is introduced for notational convenience.  We restrict ourselves to the occupation probability of the cavity state $\ket{n}\bra{n}_{\rm c}$ (see equation (\ref{eqn:RedMeasN})) and define $x(t) \equiv \bra{n}_{\rm c}\rho_{11}(t)\ket{n}_{\rm c}$.  Equation (\ref{eqn:Red1}) becomes
\begin{widetext}
\bea
\nonumber x^{(4)} &+& (\kk_2 + 2\yy_1)x^{(3)} + \Bigg[ \left( \frac{\kk_2 + \yy_1}{2}\right) \left( \frac{\kk_2 + 5\yy_1}{2}\right) + 4g^2n 
+ 4g^2 \Bigg]x^{(2)} + \left[ (\kk_2 + 2\yy_1)(2g^2 + 2g^2n) + \yy_1\left( \frac{\kk_2 + \yy_1}{2} \right)^2\right]x^{(1)} \\ 
 &+& \Bigg[ 4g^2 + 4g^4n^2 + 2g^2\yy_1\left( \frac{\kk_2 + \yy_1}{2}\right) + 8g^4n 
+ 2g^2n\yy_1\left( \frac{\kk_2 + \yy_1}{2}\right) - 4g^4(n+1)^2 \Bigg]x = 0
\label{eqn:RedOrd}
\eea
\end{widetext}
This can be further simplified upon the assumption that $\frac{1}{T_2}$ is large, as shown in equation (\ref{eqn:RedApprox}).

To obtain equation (\ref{eqn:LowT2Det}), we rewrite the equations of (\ref{eqn:FermiEqns}) in matrix form
\bea
\partial_t\left( \begin{array}{c}
P_0 \\
P_1 \\
P_m \end{array} \right) =
\left( \begin{array}{ccc}
-\yy_2n&\yy_2n&0 \\
\yy_2n&-(\yy_1 + \yy_2n)&0 \\
0&\yy_1&0 \end{array} \right)
\left( \begin{array}{c}
P_0 \\
P_1 \\
P_m \end{array} \right)
\eea
where the index indicating the photon number in the cavity has been suppressed. The Laplace transform of this system of equations is 
\bea
\label{eqn:LowT2Lap}
\left( \begin{array}{c}
s\mathcal{P}_0 - 1 \\
s\mathcal{P}_1 \\
s\mathcal{P}_m \end{array} \right) =
\left( \begin{array}{ccc}
-\yy_2n&\yy_2n&0 \\
\yy_2n&-(\yy_1 + \yy_2n)&0 \\
0&\yy_1&0 \end{array} \right)
\left( \begin{array}{c}
\mathcal{P}_0 \\
\mathcal{P}_1 \\
\mathcal{P}_m \end{array} \right)
\eea
where we have used the fact that $P_i(0) = \delta_{i0}, $ and defined $\mathcal{P}_i(s) = \mathcal{L}[P_i(t)]$.

Solving the system of equations (\ref{eqn:LowT2Lap}) for $\mathcal{P}_m(s)$ gives
\be
\label{eqn:LowT2LapDet}
\mathcal{P}_m(s) = \frac{n\yy_1\yy_2}{s\left( s^2 + s\left( 2n\yy_2 + \yy_1\right) + n\yy_1\yy_2 \right)}.
\ee
Using partial fractions and finding the residues of $\mathcal{P}_m$ at the poles allows us to rewrite equation (\ref{eqn:LowT2LapDet}) as follows:  
\be
\label{eqn:LowT2LapDetp}
\mathcal{P}_m(s) = \frac{1}{s} + \frac{\yy_1\yy_2n}{s_+ - s_-}\left(\frac{1}{s_+(s - s_+)} - \frac{1}{s_-(s - s_-)}  \right),
\ee
where $s_\pm$ are as defined in equation (\ref{eqn:LowT2roots}).
The inverse Laplace transform of equation (\ref{eqn:LowT2LapDetp}) can easily be calculated to obtain equation (\ref{eqn:LowT2Det}).

\bibliography{BABib}

\end{document}